\documentclass{article} 
\usepackage{amsmath, amssymb}
\usepackage{amsthm}
\usepackage{graphicx}
\usepackage{natbib}
\usepackage{url} 
\usepackage{hyperref}
\usepackage{booktabs}
\usepackage{xcolor, colortbl}
\usepackage{nccmath}
\usepackage{subcaption}
\usepackage{rotating}
\usepackage{multirow}
\usepackage{relsize}
\usepackage{enumitem}
\usepackage{xcolor}
\usepackage[margin=4.55cm]{geometry}
\usepackage{changes}
\definechangesauthor[name=Luca, color=red]{LM}

\newcommand{\blind}{0}

\hypersetup{
			colorlinks=true,
			linkcolor=blue,
            linktoc=page,
			anchorcolor=blue,
			citecolor=blue,
			urlcolor=blue,
 }

\def\0{\mbox{\bf{0}}}
\def\bs{\mathbf{s}}\def\be{\mathbf{e}}

\def\diag{\mbox{diag}}

\renewcommand{\baselinestretch}{1.2}
\newcommand{\argmin}{\textnormal{arg min}}
\usepackage[toc,page]{appendix}

\def \be{\begin{equation}}
\def \ee{\end{equation}}
\def \ber{\begin{eqnarray}}
\def \eer{\end{eqnarray}}
\def \berr{\begin{eqnarray*}}
\def \eerr{\end{eqnarray*}}
\def \bqmatrix{\begin{bmatrix}}
\def \eqmatrix{\end{bmatrix}}

\newtheorem{proposition}{Proposition}

\def \be{\begin{equation}}
\def \ee{\end{equation}}
\def \ber{\begin{eqnarray}}
\def \eer{\end{eqnarray}}
\def \berr{\begin{eqnarray*}}
\def \eerr{\end{eqnarray*}}
\def \bamatrix{\begin{pmatrix}}
\def \eamatrix{\end{pmatrix}}
\def \bqmatrix{\begin{bmatrix}}
\def \eqmatrix{\end{bmatrix}}

\def \argmin{{\rm argmin }}

\def \argmin{{\rm argmin }}

\def \bs{\boldsymbol}

\begin{document}

\def\spacingset#1{\renewcommand{\baselinestretch}%
{#1}\small\normalsize} \spacingset{1}


\if0\blind
{
  \title{\bf Forecasting VaR and ES using a joint quantile regression and implications in portfolio allocation \\
 }
  \author{Luca Merlo \hspace{.2cm}\\
    Department of Statistical Sciences, Sapienza University of Rome\\
    and \\
    Lea Petrella\\
    MEMOTEF Department, Sapienza University of Rome \\
    and \\
    Valentina Raponi\\
   IESE Business School, University of Navarra}
  \maketitle
} \fi

\if1\blind
{
  \bigskip
  \bigskip
  \bigskip
  \begin{center}
    {\LARGE\bf Title}
\end{center}
  \medskip
} \fi

\bigskip
\begin{abstract}
{\textcolor{black}{In this paper we propose a multivariate \textcolor{black}{quantile regression} framework to forecast Value at Risk (VaR) and Expected Shortfall (ES) of multiple financial assets simultaneously, extending \cite{taylor2017forecasting}. \textcolor{black}{We generalize} 
 the Multivariate Asymmetric Laplace (MAL) joint quantile regression of \cite{petrella2018joint} to a time-varying setting, which allows us to specify a dynamic process for \textcolor{black}{the evolution of both} VaR and ES of each asset. 
 The proposed methodology accounts for the dependence structure among asset returns. 
By exploiting the properties of the MAL distribution, we then propose a new portfolio optimization method that minimizes the portfolio risk and controls for well-known characteristics of financial data. 
\textcolor{black}{We evaluate the advantages of the proposed} approach \textcolor{black}{on both simulated and real data}, using weekly returns on three major stock market indices. \textcolor{black}{We show} that our method outperforms other existing models and provides more accurate risk measure forecasts compared to univariate ones.}  
}



\end{abstract}

\vfill

\newpage
\spacingset{1.45} 

\section{Introduction}\label{sec:intro}

The events of the ongoing credit crisis and past financial crises have emphasized the necessity for appropriate risk measures. The use of quantitative risk measures has become an essential management tool providing advice, analysis and support for financial and asset management decisions to market participants and regulators. 
The most widely used risk measure is Value at Risk (VaR). VaR measures the maximum loss which a financial operator can incur over a defined time horizon and for a given confidence level. Its clear meaning and computational ease made it very popular among practitioners, so much so that it has largely contaminated the banking regulatory framework. However, VaR has a number of drawbacks (\citealt{artzner1997thinking, artzner1999coherent}). {First, VaR} \textcolor{black}{does not
account for tail risk, i.e.} 
 it does not warn us about the size of the losses that occur with a probability lower than the predetermined confidence level. {Second, VaR}  is not a {``coherent"} risk measure (\citealt{artzner1999coherent}) since it \textcolor{black}{does not satisfy} the sub-additivity property, and hence, it does not take into consideration the benefits of diversification. {As a result, i}nvestors and risk managers are likely to construct positions with unintended weaknesses that result in greater losses under conditions beyond the VaR level (\citealt{yamai2005value}). Market participants could solve such problems by adopting the Expected Shortfall (ES) risk measure, {{which is} defined as the conditional expectation of exceedances beyond VaR  {(see \cite{acerbi2002coherence} and \cite{rockafellar2000optimization})}. {Unlike VaR, ES  is a coherent risk measure and provides} more information on the shape and the heaviness of the tails of the loss distribution. 
{Therefore, ES has gained increasing attention from risk managers, banking regulators and investors as an alternative measure of risk, complementing the VaR measure.  }

{However, despite its interesting properties, and in contrast with VaR, \textcolor{black}{little work exists} on modeling ES. This is in part due to the fact that ES is not an ``elicitable" measure, in the sense that there does not exist a loss function such that ES is the solution that minimizes the expected loss. Several works have been proposed in the literature to overcome the problem of elicitability (see, e.g., \cite{engle2004CAViaR}, \cite{Cai2008}, \cite{Taylor2008}, \cite{Zhu2011}, \cite{du2017backtesting}, \cite{Patton2019}, \cite{Bu2019}). Recently, using the results of \cite{fissler2016higher}, who show that ES is jointly elicitable with VaR, \cite{taylor2017forecasting} uses the Asymmetric Laplace (AL) distribution to jointly estimate dynamic models for both VaR and ES. In particular, \cite{taylor2017forecasting} shows that the {negative of the} log-likelihood associated to the AL distribution belongs to the class of loss functions presented in \cite{fissler2016higher}, and hence it can be used to estimate and forecast VaR and ES measures in one step. In his paper, the joint estimation of VaR and ES is obtained \textcolor{black}{in a} univariate quantile regression framework, \textcolor{black}{exploiting}
the interesting result that ES can be expressed in terms of the scale parameter of the AL density.}

\textcolor{black}{
The literature mentioned above, however, has mainly focused on univariate time series, which completely disregards the strong interrelation among assets in financial markets. \textcolor{black}{To capture the degree of tail interdependence between assets, several quantile-based methods have also been proposed to estimate VaR, without however specifying a model for the ES component;
 see, for example, the relevant works by} \cite{baur2013structure, bernardi2015bayesian, white2015var, kraus2017d} and \cite{bonaccolto2019decomposing}.}
\textcolor{black}{In this paper, we extend the univariate approach of \cite{taylor2017forecasting} to a multivariate framework, with the objective of obtaining joint estimates of both VaR and ES for multiple financial assets simultaneously, accounting for their dependence structure.  To this end, we generalize the Multivariate Asymmetric Laplace (MAL) quantile regression approach of \cite{petrella2018joint} to a time-varying setting, by allowing the parameters of the MAL to vary over time. For each asset, we model the evolution of VaR and ES as functions of the location and scale parameters of the distribution. In particular, for the VaR component, we adopt a Conditional Autoregressive Value at Risk (CAViaR) specification (\cite{engle2004CAViaR}).}


{The advantages of our methodology are manifold. Firstly, our approach is a joint modelling framework where both the model parameters and the pair (VaR, ES) of multiple returns are estimated simultaneously\textcolor{black}{, generalizing the univariate results of \cite{Bassett2004} and \cite{taylor2017forecasting}.} Secondly, our theory captures empirical characteristics of financial data such as peakedness, skewness, and heavy tails (see e.g., \cite{kraus1976skewness, friend1980co} and \cite{barone1985arbitrage}), without relying on the limitation of \textcolor{black}{normally} distributed returns. 
}


{The inferential problem is solved by developing a suitable Expectation-Maximization (EM) algorithm, {which exploits the mixture representation of the MAL distribution (see \cite{petrella2018joint}) properly generalized to the case of time-varying parameters.}} The finite sample properties of the proposed estimation method are also evaluated using a simulation exercise, where we show the validity and the robustness of our procedure under different data generating processes.


 A further contribution of the paper concerns the evaluation of VaR and ES in the context of portfolio optimization (see, e.g, \cite{Yiu2004} and \cite{Alexander2008}). \textcolor{black}{In recent years, the MAL density has attracted wide attention in the literature for its flexibility in modeling financial data (\cite{mittnik1991alternative, kotz2012laplace} and \cite{paolella2015multivariate}) and for its interesting properties that can be exploited to derive optimal portfolio allocations (see \cite{zhao2015mean} and \cite{shi2018portfolio}). \textcolor{black}{In the classic Mean-Variance (MV) methodology of \cite{markowitz1952portfolio}, portfolio risk is measured using the standard deviation of the portfolio. However, the MV approach is reasonably applicable only in cases where either returns follow a Gaussian distribution or the investors utility function is quadratic. Given the empirical evidence showing that market participants have a preference for positive skewness and they are more concerned about the downside risk (see \cite{arditti1971another} and \cite{konno1995mean} among others),} 
  the MAL distribution \textcolor{black}{could represents a more effective tool to select optimal portfolio allocations in the case of risk-averse agents.} 
} 
  \textcolor{black}{Therefore, 
  in this paper we exploit the MAL properties to incorporate skewness directly into the portfolio optimization method} 
  and to identify the optimal allocation weights. 
  We then compute the corresponding portfolio VaR and ES as a function of the multivariate structure of the data. We prove how this result follows directly from \textcolor{black}{the fact that any linear combination of the MAL components 
  is still AL distributed, with} location, skew and scale parameters that are functions of the MAL parameters and the portfolio weights. Therefore, once we obtain the Maximum Likelihood (ML) estimates of the MAL parameters from the proposed dynamic quantile regression model, we fix a desired level of risk for any target portfolio and search for the optimal allocation weights according to the adopted strategy.

 Specifically, we consider the Skewness Mean-Variance (SMV) strategy of \cite{zhao2015mean}, where the optimal allocation is obtained by minimizing the portfolio variance, while controlling for the skewness of asset returns. 
 However, \cite{zhao2015mean} employed the method of moments to estimate the portfolio variance; on the contrary, we estimate the MAL parameters in a ML framework by using an EM algorithm. 

{Empirically, we analyze  weekly returns of the FTSE 100, NIKKEI 225 and Standard $\&$ Poor's 500 (S\&P 500) market indices from April 1985 to February 2021.}
{In a first out-of-sample exercise we jointly estimate the VaR and ES of the three stock market indices using the proposed dynamic joint quantile  regression model,  hence taking into account for the correlation among the three indices.  To evaluate VaR and ES forecasts and to show the main advantages of the proposed method, we use different backtesting procedures, where we compare the out-of-sample VaR and ES predictions with the ones obtained by applying the univariate method of \cite{taylor2017forecasting}. In particular, to perform a joint evaluation of VaR and ES, we follow \cite{fissler2015expected}, \cite{nolde2017elicitability}, \cite{Patton2019} and \cite{taylor2017forecasting} and extend their approach by introducing a new scoring function based on the MAL distribution. \textcolor{black}{
 \textcolor{black}{We find that our multivariate method always provides more accurate VaR and ES predictions compared to other well-known approaches, like the Quantile AutoRegression of \cite{koenker2006quantile} and the dynamic quantile regression of \cite{taylor2017forecasting}. Moreover, in line with \cite{taylor2017forecasting}, our results show that the Asymmetric Slope CAViaR specification of \cite{engle2004CAViaR} yields the best VaR and ES forecasts for all the three indices at different quantile levels, confirming the existence of relevant asymmetries in the impact of positive and negative returns.}} 

{In a second empirical \textcolor{black}{analysis}, we aggregate the stock market indices to form a financial portfolio with a predetermined level of risk, and estimate its optimal allocation weights by implementing our new optimization procedure. We then compute the out-of-sample portfolio's VaR and ES and evaluate the predictions using the univariate backtesting procedures of \cite{taylor2017forecasting}, \cite{nolde2017elicitability} and  \cite{Patton2019}. 
 \textcolor{black}{The empirical analysis reveals that our multivariate method produces the lowest average losses compared to other existing strategies based on the multivariate Normal and t- distributions, regardless of the scoring function being used. In addition, we find that the proposed methodology overall yields the highest Sharpe Ratio and the least concentrated portfolio allocations.}
}

{The rest of the paper is organized as follows. In Section \ref{sec:meth}, we introduce the dynamic multiple quantile regression and propose a joint model for VaR and ES.  We then illustrate the EM-based ML approach for the simultaneous estimation of VaR and ES. Section \ref{sec:portfolio} develops the portfolio allocation problem. Section \ref{sec:eva} introduces a new scoring function for the joint evaluation of VaR and ES forecasts. In Section \ref{sec:emp} we discuss the main empirical results, while Section \ref{sec:conc} concludes. All the proofs are provided in Appendix A, while the simulation study is presented in Appendix B.

\section{Multivariate framework}\label{sec:meth}
\textcolor{black}{In this paper} \textcolor{black}{we generalize the univariate regression approach of \cite{taylor2017forecasting}. 
 Specifically, by extending the MAL density of \cite{petrella2018joint} --  allowing the \textcolor{black}{location and scale parameters} of the MAL to vary over time --  we  estimate the pair of VaR and ES associated to each asset 
 using a joint quantile regression framework.} \textcolor{black}{In this way, we are able to calculate the time-varying VaR and ES simultaneously for all marginal response variables, accounting for possible correlation among the considered assets.
\textcolor{black}{For the VaR components, we assume a CAViaR specification (see \cite{engle2004CAViaR}).}} Parameter estimation is carried out using a suitable EM algorithm as in \cite{petrella2018joint}, properly extended to deal with the time-varying setting. \textcolor{black}{In this way, the estimated parameters account for tail interdependence among multiple returns and convey this information onto the VaR and ES estimates.}

We start by introducing the time-varying joint quantile regression model in Section  \ref{subsec:jqreg}, where we consider a dynamic generalization of the MAL density proposed in \cite{petrella2018joint}. We then show in Section \ref{subsec:jvares} how the resulting time-varying scale parameter of the MAL can be used to model the ES vector and derive a parsimonious approach for simultaneous estimation of VaR and ES in a multidimensional setting. Parameter estimation and the EM algorithm are described in Section \ref{EM MLE}.  }

\subsection{Dynamic joint quantile regression}\label{subsec:jqreg}

{Let $\bs Y_t=[Y_{t1},Y_{t2},...,Y_{tp}]'$ be a $p$-variate response variable and denote by ${\cal Q}_{Y_{tj}}(\tau_j|\mathcal{F}_{t-1})$  the $\tau_j$-quantile function of each of the $j$-th component of $\bs Y_t$, conditional on the information set $\mathcal{F}_{t-1}$ available at time $t-1$,  for $j=1,...,p$ and $t=1,...,T$. Then,  for a given $\tau_j$, we consider the following autoregressive dynamic:

\begin{equation}\label{caviar}
{\cal Q}_{Y_{tj}}(\tau_j|\mathcal{F}_{t-1}) = \omega_j + \eta_j {\cal Q}_{Y_{t-1j}}(\tau_j|\mathcal{F}_{t-2}) + \ell ( \bs{\beta}_j, Y_{t-1j}),
\end{equation} 

\noindent where  $ \omega_j= \omega_{j} (\tau_j)$,  $\eta_j=\eta_j(\tau_j)$ and $\bs \beta_j=\bs \beta_j(\tau_j)=[\beta_{1j},..., \beta_{Kj}]'$ are model parameters that depend on the chosen level $\tau_j$ and where we suppress the index $\tau_j$ for simplicity of notation. 
The dynamic specification in (\ref{caviar}) is well-known in the literature as the CAViaR model of \cite{engle2004CAViaR}, which attempts to compute the $\tau$-th level VaR by estimating the $\tau$-th level quantile of the asset returns through a conditional autoregressive structure.
{The function $\ell(\cdot)$ \textcolor{black}{represents the so-called} News Impact Curve (NIC), originally introduced by \cite{engle1993measuring}.  For each $j$-th component, the NIC function essentially feeds back the last available observation ($ Y_{t-1j}$) into the present value of the conditional quantile, through the $K \times 1$ parameter vector $\bs{\beta}_j$.  {Following the CAViaR literature, we will consider different specifications for $\ell(\cdot)$ to model the marginal quantiles, \textcolor{black}{which will be described in the next section.}}

 Using matrix notation, the representation in (\ref{caviar}) can be embedded in the following multivariate linear regression model:

\ber \label{multivRegr}
\bs Y_t =\bs{\mu}_t + \bs \epsilon_t, \quad \quad t=1,2,...,T
\eer

\noindent  where $\bs \epsilon_t$ denotes a $p \times 1$ vector of error terms, {having} each marginal quantile (at fixed levels $\tau_1,.., \tau_p$, respectively) equal to zero,  to ensure that $\bs{\mu}_t={\cal Q}_{\bs{Y}_{t}}(\bs \tau|\mathcal{F}_{t-1})$. 

To estimate the regression model in (\ref{multivRegr}), we consider a dynamic generalization of the MAL distribution introduced in \cite{petrella2018joint} and \cite{kotz2012laplace},  {i.e. we consider the}  time-varying distribution $\mbox{MAL}_p \left(\bs{\mu}_t,  \bs D_t \bs {\tilde\xi}, \, \bs D_t \bs {\tilde\Sigma} \bs D_t  \right)$, with density function:
\ber \label{MALdensityConstr}
f_{\bs{Y}_t} (\bs y_t | \bs{\mu}_t , \bs D_t\bs {\tilde\xi}, \bs D_t\bs {\tilde\Sigma} \bs D_t, \mathcal{F}_{t-1}) =  \frac{ 2 
\exp{\left\{(\bs y_t- \bs{\mu}_t)' \bs D_t^{-1} \bs{\tilde \Sigma}^{-1}\bs {\tilde\xi} \right\}}}
{(2\pi)^{p/2} |\bs D_t \bs {\tilde \Sigma} \bs D_t|^{1/2}   } \left( \frac{\tilde m_t}{2+\tilde d}\right)^{\nu/2}K_{\nu}\left(  \sqrt{(2+\tilde d)\tilde m_t} \right).\nonumber\\[-0.1in]
\eer
In \eqref{MALdensityConstr}, $\bs{\mu}_t$ represents the location parameter vector, $\bs D_t \bs {\tilde\xi} \in {\mathcal R}^p$ is the scale (or skew) parameter, with $\bs D_t= \diag [\delta_{t1}, \delta_{t2},..., \delta_{tp}]$, $\delta_{tj}>0$ and $ \bs {\tilde\xi}= [\tilde \xi_1, \tilde \xi_2,...,\tilde  \xi_p]'$, having generic element  $\tilde \xi_j= \frac{1- 2 \tau_j}{\tau_j(1 - \tau_j)}$. $\bs {\tilde \Sigma}$ is a $p \times p$ positive definite matrix such that $\bs {\tilde \Sigma} = \bs{\tilde \Lambda} \bs \Psi \bs{\tilde \Lambda}$, with $\bs \Psi $ \textcolor{black}{having the structure of a} correlation matrix\textcolor{black}{
\footnote{\textcolor{black}{More in detail, $\mathbf{\Psi}$ represents the correlation matrix of the (latent) Gaussian process that defines the mixture representation of the MAL (see  Equation (9) in  \cite{petrella2018joint}). Moreover, by simple calculations, it is possible to show that the covariance matrix of $\bs Y$ depends on $\mathbf{\Psi}$ through the following relationship: $\bs S = \textnormal{cov} (\bs Y) = \bs D(\bs {\tilde\xi} \bs {\tilde\xi}' + \bs{ \tilde \Lambda} \bs \Psi \bs{ \tilde \Lambda}) \bs D$. In other terms, $\mathbf{\Psi}$ represents a shifted and scaled version of the sample correlation matrix of $\bs Y$ through the vector $\bs {\tilde\xi}$ and the matrix $\bs D$, respectively.}}} and $\bs{\tilde \Lambda}= \diag[\tilde \sigma_1, \tilde \sigma_1,..., \tilde \sigma_p]$, with $\tilde \sigma_j^2= \frac{2}{\tau_j (1 - \tau_j)}$, $j=1,..., p$. Moreover, $\tilde m_t= (\bs y_t- \bs{\mu}_t)' (\bs D_t \bs {\tilde \Sigma }\bs D_t)^{-1}(\bs y_t-  \bs{\mu}_t)$, $\tilde d=\bs {\tilde\xi}' \bs{ \tilde \Sigma}^{-1} \bs{\tilde \xi}$, and  $K_{\nu}(\cdot)$ denotes the modified Bessel function of the third kind with index parameter $\nu= (2-p)/2$.



Notice that, as stressed in \cite{petrella2018joint}, the specification in (\ref{MALdensityConstr}) should not be viewed as a parametric assumption in model (\ref{multivRegr}), but rather a convenient tool to jointly estimate marginal dynamic quantiles of a multivariate response variable in a quantile regression framework. 
{Moreover, as clarified in their paper, the constraints $\tilde \xi_j= \frac{1- 2 \tau_j}{\tau_j(1 - \tau_j)}$ and $\tilde \sigma_j^2= \frac{2}{\tau_j (1 - \tau_j)}$ need to be  imposed  to guarantee model identifiability (see \cite{petrella2018joint}, Proposition 2), and also to  ensure that the dynamic quantile specification in (\ref{caviar}) holds, i.e.  $\mathbb{P}(Y_{tj} < \mu_{tj}) = \tau_j$ holds for each $j=1,2,...,p$.  }

In addition, when such constraints are satisfied, then each marginal component of the MAL in \eqref{MALdensityConstr} follows a univariate AL distribution, that is $Y_{tj} \sim \textnormal{AL}(\mu_{tj}, \tau_j, \delta_{tj})$, where $\delta_{tj}$ represents the time-varying scale parameter of $Y_{tj}$. This allows us to exploit the result of  \cite{taylor2017forecasting}, who shows the link between the scale parameter of the AL distribution and the ES risk measure in a univariate framework.  By extending these results, we provide new insights on how to estimate conditional VaR and ES jointly in a multidimensional setting, \textcolor{black}{that accounts} for correlations between marginals. This is explained in details in the next section.

\subsection{Modeling VaR and ES jointly}\label{subsec:jvares}

{Following \cite{engle2004CAViaR},  the CAViaR specification in (\ref{caviar}) allows us to derive the VaR of an asset at level $\tau_j$ by estimating the corresponding quantile at the $\tau_j$-th level, through a conditional autoregressive structure. In the following, we consider several CAViaR formulations, depending on the choice of the NIC function  $\ell (\cdot)$. We then extend the idea of \cite{taylor2017forecasting} to a multivariate setting, in order to model and estimate the ES component dynamically.

The CAViaR specifications that we consider are the following:}

\ber
{\cal Q}_{Y_{tj}}(\tau_j|\mathcal{F}_{t-1}) &=& \omega_j + \eta_j {\cal Q}_{Y_{t-1j}}(\tau_j|\mathcal{F}_{t-2}) +  \beta_{1j} |Y_{t-1j}|, \quad \quad \textit{Symmetric Absolute Value (SAV)}\nonumber\\[-0.1in]
&& \label{SAV}\\
{\cal Q}_{Y_{tj}}(\tau_j|\mathcal{F}_{t-1}) &=& \omega_j + \eta_j {\cal Q}_{Y_{t-1j}}(\tau_j|\mathcal{F}_{t-2}) + \beta_{1j} Y^{+}_{t-1j} + \beta_{2j} Y^{-}_{t-1j}, \quad \quad \textit{Asymmetric Slope (AS)} \nonumber\\[-0.1in]
&& \label{AS}\\
{\cal Q}_{Y_{tj}}(\tau_j|\mathcal{F}_{t-1}) &=& \left( \omega_j + \eta_j {\cal Q}^2_{Y_{t-1j}}(\tau_j|\mathcal{F}_{t-2})  +  \beta_{2j} Y^{2}_{t-1j} \right)^{1/2} \quad \quad \textit{Indirect GARCH(1,1) (IG)}\nonumber\\[-0.1in]
&& \label{IG}
\eer

\noindent where $\bs \omega = [\omega_1, ..., \omega_p]', \bs \eta =  [\eta_1, ..., \eta_p]'$ and $\bs \beta = [\bs \beta_1, ..., \bs \beta_p]'$, with $\bs \beta_j = [\beta_{1j},\beta_{2j}]'$, are unknown parameters to be estimated, and  where $y^{+} = \max(y, 0)$ and $y^{-}= - \min(y,0)$, denote the positive and negative part of $y$, respectively. 

	}

{For the ES component, we exploit the interesting link provided in \cite{Bassett2004}, which relates univariate quantile regression to conditional ES through the following relation:

\ber \label{QESrel}
ES_{tj} = \mathbb{E}[Y_{tj}] - \frac{\mathbb{E}[(Y_{tj} - {\cal Q}_{Y_{tj}})(\tau_j - \bs{1}_{(Y_{tj} < {\cal Q}_{Y_{tj}})})]}{\tau_j}
\eer

\noindent with $\bs{1}(\cdot)$ being the indicator function. Following  \cite{taylor2017forecasting},  the expression in (\ref{QESrel}) can be rearranged so that the conditional ES can be expressed  in terms of the conditional AL scale parameter $\delta_{tj}$. Specifically, \textcolor{black}{recalling} that each marginal of the MAL distribution has a univariate AL density with conditional scale parameter equal to $\delta_{tj}={\mathbb{E}[(Y_{tj} - {\cal Q}_{Y_{tj}})(\tau_j - \bs{1}_{(Y_{tj} < {\cal Q}_{Y_{tj}})})]}$, then equation (\ref{QESrel}) reduces to:

\ber \label{ES}
ES_{tj} =\mathbb{E}[Y_{tj}]-\frac{\delta_{tj}}{\tau_j}, \quad \quad t=1,2,...,T, \quad j=1,2,...,p
\eer

\noindent implying that
\ber \label{deltaRelES}
\delta_{tj} &=& {\tau_j} \left( \mathbb{E}[Y_{tj}] -  ES_{tj} \right), \quad \quad t=1,2,...,T, \quad j=1,2,...,p
\eer

\noindent In order to ensure that each estimated ES does not cross the corresponding estimated quantile, 
we model ES in \eqref{deltaRelES} as the product of the quantile and a constant factor (see, e.g.,  \cite{gourieroux2012converting} and \cite{taylor2017forecasting}) as follows:
\ber \label{deltaES}
ES_{tj} = (1+e^{\gamma_{0j}}){\cal Q}_{Y_{tj}}(\tau_j), \quad \quad t=1,2,...,T, \quad j=1,2,...,p,
\eer

\noindent where $\gamma_{0j}$ is an unconstrained parameter to be estimated such that $1+e^{\gamma_{0j}}$ is greater than 1 and $\bs Y_t$ is assumed with zero mean. \textcolor{black}{We collect the unknown parameters in the vector $\bs{\gamma}_0 = [\gamma_{01},...,\gamma_{0p}]'$.} As explained in  \cite{taylor2017forecasting}, this formulation correctly describes the relationship between ES and VaR for different  data generating processes, such as a GARCH process
with a Student-t distribution.  Therefore, the representation in (\ref{deltaES}) provides a simple and parsimonious approach to estimate VaR and ES simultaneously in a dynamic framework.  

In \eqref{deltaES}, \textcolor{black}{however}, only the quantile is dynamic, while the factor $1+e^{\gamma_{0j}}$ remains constant through time. Therefore, in order to generalize this approach, we also consider the alternative formulation for the ES presented in \cite{taylor2017forecasting}, where the difference between the ES and the VaR is modeled using an AutoRegressive (AR) specification as follows:
\begin{align}
ES_{tj} &= {\cal Q}_{Y_{tj}}(\tau_j) - x_{tj}, \quad \quad t=1,2,...,T, \quad j=1,2,...,p, \label{ARES1}\\
x_{tj} &= (\gamma_{1j} + \gamma_{2j} ({\cal Q}_{Y_{t-1j}}(\tau_j) - Y_{t-1j}) + \gamma_{3j} x_{t-1j}) \bs{1}_{(Y_{tj} \leq {\cal Q}_{Y_{tj}})} + x_{t-1j} \bs{1}_{(Y_{tj} > {\cal Q}_{Y_{tj}})},\label{ARES2}
\end{align}
where we define the nonnegative parameter $\bs{\gamma} =[\bs \gamma_1,...,\bs \gamma_p]'$, with $\bs \gamma_j = [\gamma_{1j}, \gamma_{2j},\gamma_{3j}]'$, to ensure that the VaR and ES estimates do not cross.

\noindent In the next section, we show how to estimate model parameters using a ML approach based on a dynamic modification of the EM algorithm proposed by \cite{petrella2018joint}. 
}
{
\subsection{Parameter estimation using the EM algorithm} \label{EM MLE}

{Before describing the main steps of the EM algorithm}, { we first introduce the notation $\bs{\cal Q}_t = {\cal Q}_{\bs{Y}_{t}}(\bs \tau|\mathcal{F}_{t-1})$, $\bs D_t (\bs{ \gamma})$ and ${\bs{\tilde \Sigma}} (\bs{ \Psi})$ to make clear that the vector ${\cal Q}_{\bs{Y}_{t}}(\bs \tau|\mathcal{F}_{t-1})$ and the matrices $\bs{D}_t$ and $\bs{\tilde \Sigma}$ depend on the {unknown} parameters $\bs \omega, \bs \eta, \bs{\beta}, \bs{\gamma}$ and $\bs \Psi$, respectively.}
The derivation of the EM algorithm is based on the Proposition 3 of \cite{petrella2018joint}, properly extended to deal with the autoregressive structure of the quantile function $\bs{\cal Q}_t$ and the time dependency of the scale matrix $\bs D_t (\bs{ \gamma})$.

Let ${\bs {\Phi}} = \{\bs \omega, \bs \eta, \bs{\beta}, \bs{\gamma}, {\bs \Psi}\}$ denote the global set of parameters and define ${\bs {\hat \Phi}} = \{\bs{\hat \omega}, \bs{\hat \eta}, \bs{\hat\beta}, \bs{\hat\gamma}, {\bs {\hat\Psi}}\}$ as the corresponding set of parameter estimates. For a given vector $\bs \tau=[\tau_1, \tau_2,...,\tau_p]'$, the expected complete log-likelihood function (up to additive constants), given the observed data $\bs Y_t$ and the parameter estimates ${\bs {\hat \Phi}}$, is:

\begin{fleqn}[\parindent]
\begin{align}
E\left[\l_c({\bs {\Phi}}| \bs Y_t , \bs{\hat{\Phi}})\right] =  -\frac{1}{2} \sum_{t=1}^T \log |\bs D_{t}{({\bs \gamma})} \tilde {\bs\Sigma} {(\bs \Psi)} \bs D_{t}{({\bs \gamma})}| + \sum_{t=1}^T (\bs Y_t - \bs{\cal Q}_t)' \bs D_{t}{({\bs \gamma})}^{-1} \tilde {\bs\Sigma} {(\bs \Psi)}^{-1}\tilde {\bs \xi} \label{OF0}
\end{align}
\end{fleqn}

\begin{fleqn}[\parindent]
\begin{align}
   \quad \quad \quad & - \frac{1}{2} \sum_{t=1}^T z_t(\bs Y_t- \bs{\cal Q}_t)' (\bs D_{t}{({\bs \gamma})} \tilde  {\bs \Sigma}{({\bs \Psi})} \bs D_{t}{({\bs \gamma})})^{-1} (\bs Y_t - \bs{\cal Q}_t) \label{OF2}   \\
    & - \frac{1}{2} \tilde{{\bs \xi}}' \tilde {\bs\Sigma} {(\bs \Psi)}^{-1} \tilde {\bs \xi}\sum_{t=1}^T u_t,  \label{OF3}  
\end{align}
\end{fleqn}

\noindent where 
\ber
{u}_t &=&  E[W_t|\bs Y_t , \bs{\hat{\Phi}}]=\left(  \frac{\hat {\tilde {m}}_t}{2+\hat {\tilde d}}  \right)^{\frac{1}{2}} \frac{K_{\nu +1}\left( \sqrt{(2+\hat{\tilde d})\hat {\tilde {m}}_t} \right)}{K_{\nu}\left(   \sqrt{(2+\hat {\tilde d}) \hat {\tilde {m}}_t}\right)} \label{weightsu}\\[0.1in]
{z}_t &=&E[W_t^{-1}|\bs Y_t , \bs{\hat{\Phi}}]= \left( \frac{2+\hat {\tilde d}}{\hat {\tilde {m}}_t} \right)^{\frac{1}{2}}  \frac{K_{\nu +1}  \left( \sqrt{(2+\hat {\tilde d})\hat {\tilde {m}}_t} \right)}{K_{\nu}  \left( \sqrt{(2+\hat {\tilde d})\hat {\tilde {m}}_t} \right)} - \frac{2 \nu}{\hat {\tilde {m}}_t}  \label{weightsz},
\eer

\noindent with 
\ber
\hat {\tilde{m}}_t &=& (\bs y_t- \bs{\cal Q}_t)' (\bs D_{t}{(\hat{\bs \gamma})} \tilde {\bs\Sigma} {(\hat{\bs \Psi})} \bs D_{t}{(\hat{\bs \gamma})})^{-1}(\bs y_t- \bs{\cal Q}_t), \qquad 
\hat{ \tilde d} = \tilde {\bs \xi}' \tilde {\bs\Sigma} {(\hat{\bs \Psi})}^{-1} \tilde{ \bs \xi},
\eer

\noindent and where $W_t$ follows a standard exponential distribution.

\noindent For a given vector  $\bs \tau$, the expected complete log-likelihood in \eqref{OF0}-\eqref{OF3} is then maximized with respect to the parameter set $\bs \Phi$, yielding to the M-step updates ${\bs {\hat \Phi}} = \{\bs{\hat \omega}, \bs{\hat \eta}, \bs{\hat\beta}, \bs{\hat\gamma}, {\bs {\hat\Psi}}\}$. Notice that, unlike \cite{petrella2018joint}, closed form solutions for $\bs{\hat \omega}, \bs{\hat \eta}, \bs{\hat\beta}$ and $\bs{\hat\gamma}$ do not exist, due to the autoregressive structure of the data and, therefore, numerical optimization is required. Updated estimates of $\bs{\tilde \Sigma}(\bs{\hat \Psi})$ can be instead derived  using the following expression:

\begin{fleqn}[\parindent]
\begin{align}\label{focSigma0}
   {\tilde{ \bs \Sigma}} {(\hat{\bs{\Psi}})} & = \frac{1}{T} \sum_{t=1}^T {z}_t \bs D_{t}{(\hat{\bs \gamma})}^{-1}(\bs Y_t - \bs{\cal Q}_t) (\bs Y_t - \bs{\cal Q}_t)'\bs D_{t}{(\hat{\bs \gamma})}^{-1} \\ 
    & + \frac{1}{T}\sum_{t=1}^T {u}_t \tilde {\bs \xi} \tilde {\bs \xi}' - \frac{2}{T} \sum_{t=1}^T \bs D_{t}{(\hat{\bs \gamma})}^{-1} (\bs Y_t - \bs{\cal Q}_t) \tilde {\bs \xi}'. 
\end{align}
\end{fleqn}

{Therefore, the EM algorithm can be implemented as follows:

\noindent \textit{E-step}: Set the iteration number $h=1$. Fix the vector $\bs \tau$ at the chosen quantile levels $\tau_1,...,\tau_p$ of interest and initialize the
parameter set ${\bs {\Phi}} = \{\bs \omega, \bs \eta, \bs{\beta}, \bs{\gamma}, {\bs \Psi}\}$. 
Then, given $\bs {\hat\Phi}=\bs{\hat \Phi}^{(h)}=\{ \bs{\hat \omega}^{(h)}, \bs{\hat \eta}^{(h)}, \bs{\hat\beta}^{(h)}, \bs{\hat\gamma}^{(h)}, \bs{\hat \Psi}^{(h)} \}$, at each iteration $h$ calculate the weights:

\ber
\hat{u}^{(h)}_t &=&  \left(  \frac{\hat {\tilde {m}}^{(h)}_t}{2+\hat {\tilde d}^{(h)}}  \right)^{\frac{1}{2}} \frac{K_{\nu +1}\left( \sqrt{(2+\hat{\tilde d}^{(h)})\hat {\tilde {m}}^{(h)}_t} \right)}{K_{\nu}\left(   \sqrt{(2+\hat {\tilde d}^{(h)}) \hat {\tilde {m}}^{(h)}_t}\right)} \label{weightsuh}\\[0.1in]
\hat{z}^{(h)}_t &=& \left( \frac{2+\hat {\tilde d}^{(h)}}{\hat {\tilde {m}}^{(h)}_t} \right)^{\frac{1}{2}}  \frac{K_{\nu +1}  \left( \sqrt{(2+\hat {\tilde d}^{(h)})\hat {\tilde {m}}^{(h)}_t} \right)}{K_{\nu}  \left( \sqrt{(2+\hat {\tilde d}^{(h)})\hat {\tilde {m}}^{(h)}_t} \right)} - \frac{2 \nu}{\hat {\tilde {m}}^{(h)}_t}  \label{weightszh}
\eer
\noindent where
\ber
\hat {\tilde{m}}^{(h)}_t &=& (\bs y_t- \bs{\cal Q}_t^{(h)})' (\bs D_{t}{(\hat{\bs \gamma}^{(h)})} \tilde {\bs\Sigma} {(\hat{\bs \Psi}^{(h)})} \bs D_{t}{(\hat{\bs \gamma}^{(h)})})^{-1}(\bs y_t- \bs{\cal Q}_t^{(h)}), \\
\hat{ \tilde d}^{(h)} &=& \tilde {\bs \xi}' \tilde {\bs\Sigma} {(\hat{\bs \Psi}^{(h)})}^{-1} \tilde{ \bs \xi}.
\eer

\noindent \textit{M-step}: Use the estimates $\hat u^{(h)}_t$ and $\hat z^{(h)}_t$
 to maximize $E[ \l_c(\bs \Phi |\bs{\hat \Phi}^{(h)})]$ with respect to $\bs \Phi$, and obtain the updated  set of parameter estimates $\bs {\hat \Phi}^{(h+1)}$.
}

{The optimization procedure is iterated until convergence, that is when the difference between the likelihood function evaluated at two consecutive iterations is smaller than $10^{-5}$.} We initialize the EM algorithm by providing the univariate parameter estimates of \cite{taylor2017forecasting} on each asset, while the initial value for the correlation matrix $\bs \Psi$ in \eqref{MALdensityConstr} is calibrated using the empirical correlation matrix of the data. We fit the univariate models following the estimation procedure in \cite{engle2004CAViaR} and \cite{taylor2017forecasting}. In addition, we also consider a multiple random starts strategy with 100 different starting points to better explore the parameter space, and retain the solution corresponding to the maximum likelihood value. This strategy would prevent convergence issues and avoid the algorithm to be trapped at some local maxima. From an algorithmic point of view, the EM method exploits the Nelder-Mead and the Broyden-Fletcher-Goldfarb-Shanno (BFGS) optimization routines to obtain the \textcolor{black}{updated} estimates of $\boldsymbol{\beta}$ and $\boldsymbol{\gamma}$, meanwhile it uses \eqref{focSigma0} to compute the \textcolor{black}{updated} estimate of $\bs{\tilde{\Sigma}}$.
\\
The computational analysis has been conducted using the \texttt{R} (version 4.0.2) software where the functions to update $\boldsymbol{\beta}$, $\boldsymbol{\gamma}$ and $\bs{\tilde{\Sigma}}$ were coded in efficient C\texttt{++} object-oriented programming.
\\
The validity and the performance of the proposed EM algorithm have been assessed using also a simulation exercise (see Appendix B).


} 

{
\section{Portfolio Construction}

\label{sec:portfolio}
\textcolor{black}{In this section we approach the problem of portfolio allocation. Particularly, we construct the Skewness Mean-Variance (SMV) portfolio of \cite{zhao2015mean}, taking into account both for the multivariate structure and the skewness of asset returns. Following \cite{stolfi2018sparse} and \cite{zhao2015mean}, we exploit an interesting property characterizing the MAL distribution in \eqref{MALdensityConstr}. Specifically, we show that any linear combination of its marginal components follows a univariate AL distribution, whose parameters are a function of the MAL parameters in \eqref{MALdensityConstr}.} 
 \textcolor{black}{Note that, while the MAL density has so far been regarded as a convenient tool for estimating the marginal quantiles, in this section the MAL distribution is used as a data driven assumption to describe the empirical characteristics of asset returns. As already stated in the Introduction, this choice has been positively accepted in the recent financial literature to detect peakedness, fat-tail, and skewness of financial assets, overcoming possible deficiencies of standard approaches relying, for example, on the Gaussian distribution assumption.} We then evaluate the riskiness of the selected portfolio by calculating its corresponding VaR and ES, using the results of Section \ref{subsec:jvares}.

\subsection{Linear combinations of MAL {components}}\label{subsec:pdist}
{
Let assume that $\bs Y_t$ is a $p$-dimensional random variable describing the joint dynamics of $p$ \textcolor{black}{variables} at time $t$. \textcolor{black}{Let us consider a linear combination (with weights to be determined) of each component of $\bs Y_t$.} 
 \textcolor{black}{Then, the following proposition holds.} 
\begin{proposition} \label{ALDuni}
Let $\bs Y_t \sim \mbox{MAL}_p(\bs \mu_t, \bs D_t \bs{\tilde  \xi}, \bs D_t \bs {\tilde\Sigma} \bs D_t) $, with density function defined in (\ref{MALdensityConstr}). Let  $\bs b_t=(b_{t1},...,b_{tp})' \in {\cal R}^p$ be a vector of weights such that  $\bs b_t \neq \bs{0}_p$, with $ \bs{0}_p$ denoting  a $p$-vector of zeros. Define the random variable $Y_t^{\bs b} = \sum_{j=1}^p b_{tj} Y_{tj}$. Then,  
\ber
Y_t^{\bs b}  \sim \mbox{AL} (\mu_t^\star, \tau_t^\star, \delta_t^\star)
\eer

\noindent where

\ber \label{pars}
\mu_t^\star = \bs{b_t'\mu_t} \mbox{,} \quad  \tau_t^\star = \frac{1}{2}(1-\frac{\bs{b_t'D_t\tilde\xi}}{\sqrt{2(\bs{b_t'D_t\tilde\Sigma D_tb_t}) + (\bs{b_t'D_t\tilde\xi})^2}})  \quad  \textnormal{and}  \quad \delta_t^\star=\frac{(\bs{b_t'D_t\tilde\Sigma D_tb_t})}{2{\sqrt{2(\bs{b_t'D_t\tilde\Sigma D_tb_t}) + (\bs{b_t'D_t\tilde\xi})^2}}}. \nonumber \\[-0.01in]
\eer 

\end{proposition}

Proposition \ref{ALDuni} brings out two main considerations. First, the distribution of $Y_t^{\bs b}$ is still AL, which greatly facilitates the computation of the VaR and the ES in our context. Second,  the parameters of  $Y_t^{\bs b}$  are expressed as a function of the multivariate parameters $\bs \mu_t, \bs D_t$ and $\bs{\tilde \Sigma}$ of the MAL in \eqref{MALdensityConstr}. This allows us to take into account  
 the possible association among the marginal components of $ {\bs Y_t}$ when choosing the allocation weights $\bs b_t$. In the next section, we exploit such property to retrieve the returns distribution of a financial portfolio, whose optimal weights can be derived by solving a simple constrained optimization problem.  Given the resulting optimal portfolio weights,  we can then  use the results of Section \ref{subsec:jvares}   to derive  appropriate measures of portfolio's VaR and ES.} 


{\subsection{{The portfolio optimization problem}}\label{subsec:pfopt}

{ Assume that 
$\bs Y_t$ follows the distribution in \eqref{MALdensityConstr}. 
Given such specification, at each time $t$, investors might be interested in deriving a portfolio $Y^b_t= \sum_{j=1}^p b_{tj}Y_{tj}$, by investing a portion $b_{tj}$ of their capital on the asset $Y_{tj}$,   such that $\sum_{j=1}^p b_{tj}=1$. Then, under this setting, the result of Proposition \ref{ALDuni} can be easily applied, yielding a portfolio with location, skewness and scale parameters equal to, respectively, $\mu^\star_t$, $\tau^\star_t$ and  
$\delta_t^\star$ as in (\ref{pars}).

}
{Typically, in risk management applications, the skewness parameter is fixed a priori by the researcher at a certain level (constant over time) $\tau^\star_t= \tilde \tau$, as it essentially measures the overall riskiness of a financial product (a portfolio, in our case).}
Therefore, once we estimate the time-varying MAL parameters from the quantile regression model in \eqref{multivRegr}, for a fixed level of risk $\tilde \tau$, the investor's portfolio decision is based on the solution of the selected portfolio strategy. \textcolor{black}{As stated above, to obtain the optimal portfolio allocation, we adopt the SMV strategy of \cite{zhao2015mean}, which seeks to minimize the portfolio variance and at the same time control for the skewness of asset returns. Formally, according to Proposition \ref{ALDuni}, the SMV portfolio solves the following constrained optimization problem:
\begin{subequations}
\begin{align}
\underset{\bs{b}_t \in \mathcal{R}^p}{\argmin} & \quad \bs{b}_t' \bs{D}_t \bs{\tilde \Sigma} \bs{D}_t \bs{b}_t & \label{obj}\\
\text{s.t.} & \quad \tau_t^\star = \tilde \tau, \quad \forall t  \label{con} \\ 
& \quad \bs{b}'_t \bs{1}_p = 1
\end{align}
\end{subequations}
where $\bs{ \tilde \Sigma}$ has been introduced in \eqref{MALdensityConstr} and accounts for the covariance matrix of the returns}, while $\bs{b}_t$ denotes the portfolio's weights at time $t$ held by the investor over the period $[t, t + 1)$. 

From an empirical point of view, the constraint in \eqref{con} implies that the portfolio weights must be adjusted at each holding period to guarantee that the VaR of the portfolio has constant {level} $\tilde \tau$, namely $\mathbb{P}(Y^{\bs{b}}_t < \mu^{\star}_t | \mathcal{F}_{t-1}) = \tilde \tau$. 
{Once we obtain the optimal portfolio weights for the period $[t, t + 1)$, we can then compute the conditional portfolio's VaR and ES  at level $\tilde \tau$ by simply applying the result  in \eqref{ES} to the univariate case. 

\noindent{As already explained above, since the parameters $\mu_t^\star, \tau_t^\star, \delta_t^\star$ depend on the parameter estimates of the MAL, information on the dependence structure \textcolor{black}{and on the empirical characteristics} embedded in the data is channeled through such estimates into the portfolio's VaR and ES forecasts.
This motivates our approach even further, since it could offer  an operative and useful tool to help investors and asset managers in deriving optimal portfolio allocations and, at the same time,  monitoring multiple VaR and ES jointly.

{
\section{Assessment of VaR and ES forecasts}\label{sec:eva}

 To assess the performance of VaR and ES predictions jointly, we introduce a new backtesting procedure, based on the multivariate approach discussed in Section  \ref{sec:meth}.
 
Backtesting techniques \textcolor{black}{are based} on quantitative tests which  scrutinize  model performance in terms of accuracy and \textcolor{black}{precision} with respect to a defined criterion.
Existing approaches, however, rely on tests that analyze VaR and ES predictions separately, i.e. they only focus on the individual evaluation of one risk measure or the other.
 VaR  evaluation is typically based on coverage tests, which measure the percentage of times that the returns have exceeded the estimated VaR at a chosen probability level $\tau$ (see, e.g., the unconditional coverage (LR$_{uc}$) test of \cite{kupiec1995techniques}, the conditional coverage (LR$_{cc}$) test of \cite{christoffersen1998evaluating} and the Dynamic Quantile (DQ) test of \cite{engle2004CAViaR}).

 \textcolor{black}{To evaluate ES forecasts, the backtesting analysis becomes more complicated} since ES is not an elicitable measure \citep{gneiting2011making} and, therefore, suitable scoring functions cannot be determined \citep{taylor2017forecasting}.  The test of \cite{mcneil2000estimation}  is \textcolor{black}{commonly} used in this context, which is based on  the discrepancy between the observed return and the ES forecast for the periods in which the return exceeds the VaR forecast. \textcolor{black}{Another suitable option is the backtesting procedure of \cite{du2017backtesting} which is based on the Unconditional ES (U$_{ES}$) and Conditional ES (C$_{ES}$) tests.} 

However, since ES relies on observations exceeding the VaR, it is clear that assessment of ES forecasts cannot be independent of the predicted VaR values. 
This, together with the fact that ES is not elicitable, motivates the introduction of a scoring function for jointly evaluating VaR and ES forecasts. Based on the characterization of consistent scoring functions introduced by \cite{fissler2016higher} and \cite{nolde2017elicitability}, several scoring rules have been proposed in the literature for the univariate setting (see, e.g., \cite{Patton2019}, \cite{fissler2015expected} and \cite{taylor2017forecasting}).  

In the following, we provide a new scoring rule that can be used  in a multivariate setting, to jointly evaluate VaR and ES forecasts of multiple (and possibly correlated) financial assets. To provide support for \textcolor{black}{our proposal of} estimating multiple VaR and ES by maximizing the MAL likelihood, we define a new scoring function ($S_{MAL}$) using the negative of the MAL log score:

\ber \label{lossmal}
\begin{aligned}
S_{MAL}\left(\bs{\cal Q}_{t}, \bs{ES}_t, \bs{y}_t; \bs {\tilde \Sigma}, \bs{\tau}\right) = \frac{1}{2} \log \left( | \bs {\tilde \Sigma} | \right) + \log \left( |(\bs{\tau} \bs{ES}_t ') \circ \textnormal{\textbf{I}}_p| \right) - \frac{\nu}{2} \log \left(  \frac{\tilde m_t}{2+ \tilde d} \right) \\ + (\bs y_t- \bs{\cal Q}_t)' {\left( (\bs{\tau} \bs{ES}_t ') \circ \textnormal{\textbf{I}}_p \right)}^{-1}  \bs{\tilde \Sigma}^{-1}\bs {\tilde\xi} - \log \left(  K_{\nu}\left( \sqrt{(2+\tilde d)\tilde m_t} \right) \right)
\end{aligned}
\eer
where $\circ$ denotes the Hadamard product and $\textnormal{\textbf{I}}_p$ represents the identity matrix of order $p$.

Notice that, when $p$ is equal to 1, the $S_{MAL}$ in \eqref{lossmal} reduces to the AL log score of \cite{taylor2017forecasting}. When $p>1$, the loss function $S_{MAL}$ allows us \textcolor{black}{to: (i) perform a joint assessment of the pairs (VaR, ES) specific to each asset and, at the same time, (ii) control for the existing correlation among returns.}

\section{Empirical Study}\label{sec:emp}
In this section we apply the methodology presented in Sections \ref{sec:meth} and \ref{sec:portfolio} to real data in order to evaluate and compare its empirical implications with the ones obtained by using a univariate framework. Particularly, we follow \cite{taylor2017forecasting} and use weekly returns of the FTSE 100, NIKKEI 225, and S\&P 500 stock market indices, from April 26, 1985 to February 01, 2021. Using a rolling window exercise, we estimate the one-week-ahead VaR and ES forecasts implied by the CAViaR specifications described in Section \ref{subsec:jvares}, and select the most desirable model  using  the \cite{diebold2002comparing} test.
In a second empirical exercise,  we aggregate the market indices to form a financial portfolio and determine  its optimal allocation weights  by solving the optimization problem described in Section \ref{subsec:pfopt}. We finally compute and assess the resulting portfolio's conditional VaR and ES for the out-of-sample period, which consists of the last 368 observations of the sample.

\subsection{Data}\label{subsec:data}
Our sample is collected from Bloomberg and it consists of 1868 weekly returns for each of the three stock indices. The main summary statistics are displayed below in Table \ref{tab:stat}, providing evidence of the well-known stylized facts on fat tails, high kurtosis, serial and cross-sectional correlation that typically characterize financial assets. Moreover, all series exhibit a negative skewness, the Jarque-Bera test significantly rejects the normality assumption, the Ljung-Box test advocates the presence of serial correlation and the Augmented Dickey-Fuller test supports the hypothesis of absence of unit roots. These results clearly motivate us to consider a quantile regression approach as investigative tool.

\subsection{Out-of-sample VaR and ES forecasting}\label{subsec:fore}



Using the approach introduced in Section \ref{sec:meth}, in this section we derive a joint estimation of VaR and ES for the three stock market indices described above. 
Particularly, we estimate the out-of-sample series of VaR and ES by considering the three different specifications in (\ref{SAV}), \eqref{AS} and \eqref{IG}, with both the multiplicative factor in \eqref{deltaES} and the AR formulation in \eqref{ARES1}-\eqref{ARES2} for the ES component. Moreover, since we are concerned with the downside risk, we evaluate the out-of-sample forecasts at three different probability levels, namely $\bs {\tau} = [0.1, 0.1, 0.1], \bs {\tau} = [0.05, 0.05, 0.05]$ and $\bs {\tau} = [0.01, 0.01, 0.01]$.

The first objective is to assess the performance of the CAViaR specifications using the proposed multivariate framework. We start by evaluating the VaR forecasts using the conventional LR$_{uc}$, LR$_{cc}$ and the DQ tests, \textcolor{black}{ while we perform the U$_{ES}$ and C$_{ES}$ tests of \cite{du2017backtesting} to evaluate the ES predictions.} The results are shown in Table \ref{tab:back}, where Panel A refers to the case of the ES vector modeled as in \eqref{deltaES}, while Panel B refers to the AR specification in \eqref{ARES1}-\eqref{ARES2}. \textcolor{black}{Looking at VaR forecasts,} in both panels, for all the three indices and for all the three quantile levels, we find that the CAViaR-AS specification is always successfully backtested at the 5\% significance level. The results are less clear, instead, for the other two CAViaR specifications. \textcolor{black}{The same results are confirmed when evaluating the ES predictions, as the CAViaR-AS specification yields again outstanding performances for all the three indices and for all the three quantile levels.} 
\\
To jointly evaluate VaR and ES forecasts associated to each stock market index, in addition to the results of the coverage tests, Table \ref{tab:back_score} reports the values of the loss functions $S_{FZN}$ of \cite{nolde2017elicitability} and $S_{FZ0}$ of \cite{Patton2019}, averaged over the out-of-sample period where:
\ber \label{lossfzn}
S_{FZN}({\cal Q}_{t}, ES_t, y_t) = (\mathbf{1}_{(y_t < {\cal Q}_{t})} - \tau) \frac{{\cal Q}_{t}}{2 \tau \sqrt{- ES_t}} - \frac{1}{2 \sqrt{- ES_t}} (\mathbf{1}_{(y_t < {\cal Q}_{t})} \frac{y_t}{\tau} - ES_t) + \sqrt{- ES_t}
\eer
and
\ber \label{lossfzg}
S_{FZ0}({\cal Q}_{t}, ES_t, y_t) = \frac{1}{\tau ES_t} \mathbf{1}_{(y_t < {\cal Q}_{t})} (y_t - {\cal Q}_{t}) + \frac{{\cal Q}_{t}}{ES_t} + \log (-ES_t) - 1.
\eer
The losses in (\ref{lossfzn}) and (\ref{lossfzg}) belong to the class of scoring rules proposed in \cite{nolde2017elicitability} and \cite{Patton2019} and have the additional advantage of generating loss differences (between competing forecasts) that are homogeneous of degree $1/2$ and zero, respectively.
\\
Overall, the results show that both the CAViaR-AS and CAViaR-IG dynamics are associated with smaller losses compared to the CAViaR-SAV model, except for the case of $\bs {\tau} = [0.1, 0.1, 0.1]$. Moreover, in line with \cite{taylor2017forecasting}, we find evidence of a better forecasting performance when using the constant multiplicative factor $(1 + e^{\gamma_{0j}})$ to model the ES parameter (Panel A), compared to the AR dynamics (Panel B).
\\
Finally, to reinforce our analysis, we evaluate the forecasting performance of the three CAViaR competing models using the scoring function in \eqref{lossmal}. Specifically, at each time $t$, and for the specified level  $\bs {\tau}$, we define by  $S_{MAL_{t}}^{(j)}(\bs{\tau})$ the scoring function associated to model $j$,  and denote the difference between the scoring function of model $i$ and model $j$ by $\Delta_{MAL, t}^{(i,j)} =S_{MAL_{t}}^{(i)}(\bs{\tau}) - S_{MAL_{t}}^{(j)}(\bs{\tau})$, where $i,j=1,2,3$. We then test for the null hypothesis that  $\mathbb{E}[\Delta_{MAL, t}^{(i,j)}]=0$ against $\mathbb{E}[\Delta_{MAL, t}^{(i,j)}]<0$ using the \cite{diebold2002comparing} test, for all the pairs of models $i$ and $j$. 
If the null hypothesis is rejected, then forecasts delivered by model $i$ are more accurate than those of model $j$, and therefore model $i$ is preferable than model $j$. The results of the test, together with the corresponding p-values, are reported in Table \ref{tab:dieboldMAL}. The table clearly shows that the CAViaR-AS specification outperforms both the CAViaR-IG and CAViaR-SAV models at all the three quantile levels and for both the adopted ES formulations of constant multiplicative factor (Panel A) and AR dynamics (Panel B). Therefore, in order to select the best performing model between the two CAViaR-AS in Panels A and B, we apply again the \cite{diebold2002comparing} test on the two competing CAViaR-AS specifications. The results are reported in Table \ref{tab:diebold2} and suggest that the CAViaR-AS model with the ES specified as a constant multiple of the VaR provides the most accurate predictions at all the three quantile levels. This is in line with \cite{taylor2017forecasting}, who also find that the same specification not only produces the smallest losses, but it also delivers the most accurate predictions than all the other competing CAViaR dynamics\footnote{To further justify this choice, we also compare the CAViaR-AS model with the Quantile AutoRegression of \cite{koenker2006quantile}. Specifically, we estimate the regression model of \cite{petrella2018joint} using the lagged returns (at lag 1) of each asset as covariates.  Comparing these two models would allow one to evaluate the potential contribution of assuming a CAViaR specification in the quantile dynamics. According to the coverage tests and the scoring functions defined in \eqref{lossfzn} and \eqref{lossfzg}, we still find a better performance of the CAViaR-AS specifications.}. These results corroborate the fact that accounting for asymmetries in the autoregressive process of a given quantile improves the model's forecasting ability (see, e.g., \cite{engle2004CAViaR}, \cite{xiliang2009estimation}, \cite{taylor2005generating} and \cite{laporta2018selection}).

 To show the advantages and the different implications of our approach, \textcolor{black}{we compare our results with the ones obtained by considering each asset separately, as if we ignored their possible dependence structure. Specifically, the three CAViaR models are estimated individually for each stock market index using the univariate approach of \cite{taylor2017forecasting}. To assess the performance of the three models} and to combine the individual forecasts of the three indices in a single value, we use the sum of the three corresponding AL log scores (see \cite{taylor2017forecasting}) as consistent scoring rule.  That is, at each time $t$, and for each model $j$, we define the following scoring function:

\ber \label{lossAL}
S_{AL_{t}}^{(j)}(\bs{\tau}) = \sum_{p=1}^3 S_{AL_{p,t}}^{(j)}(\tau_p)
\eer 

\noindent where $S_{AL_{p,t}}^{(j)}(\tau_p)$ denotes the AL log-score of  \cite{taylor2017forecasting},  corresponding to model $j$ and asset $p$: 

\ber \label{lossALu}
S_{AL_{p,t}}^{(j)}(\tau_p)= -\log \left(\frac{\tau_p-1}{ES^{(j)}_{p,t}}\right) - \frac{\left(y_{p,t} - {\cal Q}_{p,t}^{(j)}\right)\left(\tau_p - \mathbf{1}_{(y_{p,t} < {\cal Q}_{p,t}^{(j)})}\right)}{\tau_p ES_{p,t}^{(j)}}.
\eer

\noindent As explained in \cite{frongillo2015vector}, summing up the three AL scoring functions would produce a consistent scoring rule in this case, since each function $S_{AL_{p,t}}^{(j)}(\tau_p)$ elicits the pair (VaR, ES) for the corresponding $p$-th asset (see \cite{fissler2016higher} and \cite{taylor2017forecasting}). 

\noindent Then, as before, we define the difference between the scoring function of model $i$ and model $j$ by $\Delta_{AL, t}^{(i,j)} =S_{AL_{t}}^{(i)}(\bs{\tau}) - S_{AL_{t}}^{(j)}(\bs{\tau})$ and apply the \cite{diebold2002comparing} test to look for the best model in terms of forecasting accuracy. The results are reported in Table \ref{tab:dieboldAL} (Panels A and B). As shown in the table, the conclusion of the test is now less clear and does not provide any significant evidence in favor of a particular model. This is one of the first advantages of our approach, as we are able to identify a clear hierarchy among competing models. 
 

A second question of interest concerns the ``efficiency gain" of the multiple approach compared to the univariate one. In this sense, we would like to test whether taking into account the association structure among the market indices would provide us with better predictions in terms of VaR and ES. To do that, we use the backtesting procedure to identify the most ``efficient'' model, that is, the model producing the best forecasts according to the \cite{diebold2002comparing} test. To measure the efficiency gain we analyze the difference, if any, in the predictive accuracy between the forecast (VaR, ES) produced by our multivariate approach and the univariate ones.
Therefore, for a given CAViaR specification, we test for the difference between the scores obtained with the scoring rule in \eqref{lossAL} and the ones obtained with \eqref{lossmal}. The null hypothesis is that, on average, the difference is not statistically different from zero, i.e. the two approaches have the same forecasting performance. The alternative hypothesis is that the difference is smaller than zero, i.e. the multivariate approach delivers significantly better predictions (smaller losses). 

Table \ref{tab:dieboldmarg} shows the resulting test statistics and the corresponding p-values for each of the possible pairs of competing models. Interestingly, for all the considered risk levels and for all the three CAViaR specifications, we are always able to reject the null hypothesis at 5\% level, providing evidence of the efficiency gain of our proposed joint approach\footnote{As a further robustness check, we have also considered the best CAViaR specification for each asset and then applied the \cite{diebold2002comparing} test. The proposed joint method still appears to be more efficient than the univariate one of \cite{taylor2017forecasting}, making our findings unchanged.}.

To offer a graphical intuition of the results, Figure \ref{diff} shows the time series of the difference between the scoring function in  \eqref{lossAL} consistent with the univariate approach and the scoring function proposed in \eqref{lossmal}, over the whole out-of-sample period and for the three considered CAViaR specifications with the ES modeled as a multiple of VaR. The left graph in Figure \ref{diff} refers to the case of $\bs {\tau} = [0.1, 0.1, 0.1]$, the center graph displays the case of $\bs {\tau} = [0.05, 0.05, 0.05]$, while the right plot is for $\bs {\tau} = [0.01, 0.01, 0.01]$. The black line represents the difference of the two scoring functions obtained by using the CAViaR-SAV specification in \eqref{SAV}, while the red and the blue lines refer to the CAViaR-AS and CAViaR-IG dynamics, respectively.  The efficiency gain of the multivariate approach clearly emerges from these pictures. Indeed, for all the considered risk levels, and regardless of the dynamic specification of the quantiles, the difference between the two approaches is almost always positive. \textcolor{black}{This confirms} the idea that the losses associated with the univariate model can be very large if the dependence structure of the data is not accounted for.   

In Figure \ref{VARandES}, we display the series of the out-of-sample forecasts of VaR and ES, for each of the three stock indices, estimated by assuming the selected CAViaR-AS specification in both the univariate and the joint approaches. VaR predictions obtained with the univariate approach of \cite{taylor2017forecasting} are represented by the dotted blue line, while the VaR estimates produced by our joint approach are depicted in the solid red line. The estimated ES is represented by the dotted green line (using the univariate approach) and the solid orange line (using the joint approach). Left panels of Figure \ref{VARandES} refer to $\bs {\tau} = [0.1, 0.1, 0.1]$, the center panels report the case $\bs {\tau} = [0.05, 0.05, 0.05]$ while the right ones consider $\bs {\tau} = [0.01, 0.01, 0.01]$. The gray dots denote the original return series of each stock index. In all the cases, the estimates of VaR and ES produced by our joint approach lie below the corresponding values obtained with the univariate setting, suggesting that our proposed method could lead to more conservative results.

 Finally, to get a more intuitive representation of the relationship between the estimated VaR and the ES over time and across the quantile levels, Figure \ref{VARandES2} displays the absolute difference between the out-of-sample VaR and ES forecasts for each of the three stock indices. The plots in the first row are obtained by assuming the CAViaR-AS specification with the ES modeled as in \eqref{deltaES}, while the plots in the second row refer to the case of the CAViaR-AS specification with the ES following the dynamic in \eqref{ARES1}-\eqref{ARES2}, at the $\bs{\tau} = [0.1, 0.1, 0.1]$ (left column), $\bs{\tau} = [0.05, 0.05, 0.05]$ (center column) and $\bs{\tau} = [0.01, 0.01, 0.01]$ (right column) quantile levels. The blue, red and orange lines refer to the FTSE 100, NIKKEI 225 and S\&P 500 stock market indices, respectively, while the grey bands correspond to the main recession periods and to various economic and financial crises occurred since 2014. As one could reasonably expect, such difference follows the overall market volatility. Indeed, we find that the difference between the estimated risk measures is typically smaller in calm periods and larger in period of turbulent markets, with more pronounced upward spikes when the AR dynamics for the ES is used. High volatility is also clearly evident in correspondence and in the aftermath of major economic and financial crises, such as, for example, the Chinese stock market crash at the start of 2016, the Brexit in 2018 and the outbreak of COVID-19 pandemic in 2020.
  
 Based on these considerations, in the next section we consider the CAViaR-AS specification in \eqref{AS} with the ES expressed as in \eqref{deltaES}, to implement the portfolio optimization procedure.

\subsection{Out-of-sample portfolio VaR and ES forecasting}\label{subsec:port}
\textcolor{black}{In this section we use the three stock market indices of FTSE 100, NIKKEI 225 and S\&P 500 to build a SMV portfolio that delivers a certain fixed level of risk $\tilde \tau$. The optimal allocation weights are determined by solving the optimization problem described in Section \ref{subsec:pfopt}, using the parameter estimates provided by the CAViaR-AS specification in Section \ref{subsec:fore}}. 

 We evaluate the benefits of our approach by considering alternative strategies. First, we use the estimation method of \cite{zhao2015mean}, where the covariance matrix $\bs{ \tilde \Sigma}$ in \eqref{obj} is estimated using the sample variance and the sample mean of the return series. We call this strategy Moment-SMV. Second, we evaluate the classic MV of \cite{markowitz1952portfolio}. In this case, we model the conditional covariance of asset returns using several well-known autoregressive dynamics, i.e. the multivariate GARCH Dynamic Conditional Correlation model (\citealt{engle2002dynamic}), under both the multivariate Normal (MV-G-DCC-N) and Student-t (MV-G-DCC-t) error distributions, and the asymmetric Dynamic Conditional Correlation model with multivariate Normal (MV-G-aDCC-N) and Student-t (MV-G-aDCC-t) errors. Moreover, since the MV strategy could often be inadequate in controlling for asymmetric risk-averse agents, we also consider the above strategies under the multivariate skew Normal (SN) and the multivariate skew Student-t (St) distributions of \cite{bauwens2005new} as further competing strategies, which we denote by  MV-G-DCC-SN, MV-G-DCC-St, MV-G-aDCC-SN and MV-G-aDCC-St, respectively. Then, for each model we forecast the one-week-ahead conditional covariance matrix and plug it into the portfolio optimization problem.

 We \textcolor{black}{jointly estimate} VaR and ES of the resulting portfolios and analyze their out-of-sample performance using the last 368 returns of the sample. The backtesting results for the considered strategies are shown in Table \ref{tab:pfback}, where we report the AL log-score of \cite{taylor2017forecasting}, together with the S$_{FZN}$ and S$_{FZ0}$ loss functions in \eqref{lossfzn} and \eqref{lossfzg} for the joint evaluation of the pair (VaR, ES). The results clearly make our approach stand out compared to the other strategies. Indeed, \textcolor{black}{the strategies based on the multivariate Normal and t- distributions, and their skewed counterparts,} produce highly volatile VaR forecasts and suffer from larger average losses over the out-of-sample period. On the other hand, the SMV and Moment-SMV models deliver better performance gains over the MV portfolios, with the SMV being preferred at the three VaR levels, especially at the most extreme case of $\tau = 0.01$. Such gain may be traced back to the higher efficiency in the estimation procedure based on the ML approach proposed in Section \ref{EM MLE}. It is worth noticing that these conclusions remain still valid even when we use the $S_{FZ0}$ and the $S_{FZN}$ scoring rules, which do not directly depend on the AL likelihood function.} 
 
 From a financial viewpoint, in Table \ref{tab:pfback} we also evaluate the risk-adjusted returns of the competing portfolios, measured by the Sharpe Ratio (SR) and the Herfindahl-Hirchman Index of weights concentration (HHI). We find that the SMV strategy delivers the portfolio with the highest SR and the least concentrated portfolios at both $\tau = 0.1$ and $\tau = 0.05$, with an average HHI of $0.558$ and $0.557$, respectively. On the other hand, when $\tau=0.01$, the MV strategy seems to yield the portfolios with the highest SR, while the Mom-SMV strategy produces the lowest degree of weights concentration.
 
  A graphical representation of the SMV portfolio weights and their evolution over time  is provided in Figure \ref{portfolioweights}. In each plot of the figure, the blue line denotes the allocation weights assigned to the FTSE 100, the red line refers to the NIKKEI 225, while the allocation weights of the S\&P 500 are displayed in orange. The left panel considers $\tau=0.1$, the center one $\tau=0.05$, while in the right-hand side we plot the results for $\tau=0.01$. As one can see, the SMV strategy tends to invest mainly in the FTSE 100 and in the S\&P 500, while it always tends to hold only a small short position on the NIKKEI 225. It is also interesting to notice that the portfolio weights exhibit the highest volatility in periods of high uncertainty in the market, such as, for example at the end of 2016 and during the ongoing COVID-19 pandemic. Finally, we evaluate the evolution of the wealth generated by the portfolios at the three risk levels during the out-of-sample period. Figure \ref{compound} highlights a positive trend for all quantile levels $\tau=0.1$ (violet line), $\tau=0.05$ (green line) and $\tau=0.01$ (yellow line) from 2014 to 2015 and from 2016 until the outbreak of COVID-19 at the beginning of 2020.

\section{Discussion and conclusions}\label{sec:conc}
This paper proposes a dynamic joint quantile regression model for estimating VaR and ES of multiple financial assets in one step, extending the univariate framework of \cite{taylor2017forecasting}. 
To implement the methodology, we suggest a likelihood-based approach based on the MAL density proposed in \cite{petrella2018joint}, generalized to the case of time-varying parameters. This offers a powerful tool to model the dynamics of  multiple  VaR and ES jointly. \textcolor{black}{Indeed, the location parameter of the MAL density represents the vector of VaRs, while the ES can be expressed as a simple function of the density scale parameter.}   

We show that our approach can offer several important advantages, both theoretical and practical. First, it provides a significant gain in terms of estimation efficiency, as it allows us to estimate multiple VaR and ES in just one step. Second, it can significantly improve \textcolor{black}{forecasts accuracy}, since it accounts for the dependence structure among financial assets, that cannot be detected by univariate methods.   
These results are also confirmed empirically. Indeed, using three stock
market indices as in \cite{taylor2017forecasting}, we  estimate the pairs of (VaR, ES) for each of the three assets and evaluate the forecasts using a new scoring function based on the  MAL density, which allows us to account  for the dependence structure among the considered assets at each point in time. The forecasts of VaR and ES have been compared with the ones obtained by the univariate approach of \cite{taylor2017forecasting}, i.e. by considering the three stock market indices separately, as they were independent to each other. What we find is a significant gain in terms of forecasting accuracy using the proposed multivariate framework,  leading to more reliable risk measure estimates.

\textcolor{black}{Following \cite{zhao2015mean},} we also exploit the properties of the time-varying MAL distribution to derive a new portfolio optimization method, where the optimal allocation weights are adjusted at each holding period to ensure that the portfolio meets a predetermined level of risk. \textcolor{black}{Empirically, we find that our optimization method produces a portfolio with less concentrated allocation weights and higher Sharpe Ratio compared to other existing strategies.}

Several extensions and generalizations could be analyzed, leaving space for future research. Although we have only focused on CAViaR models, one could consider other VaR based models in the quantile regression framework and/or specify different ES dynamics where the factor $(1 + e^{\gamma_{tj}})$ varies over time according to an autoregressive process for $\gamma_{tj}$. Another interesting research problem would involve the evaluation of the portfolio performance when a larger set of indices is considered, which may help us in providing an empirical ranking based on their VaR and ES forecasts.
 In this case, a penalized approach, as done, for instance, by \cite{petrella2018joint}, could be adopted to deal with the curse of dimensionality, improve estimation, gain in parsimony and conduct a variable selection procedure. Finally, other portfolio strategies can be implemented  as well, where the choice of the weights might be motivated by other practical considerations or regulatory restrictions.

\section*{Acknowledgements}
This project has received funding from the postdoctoral fellowships programme Beatriu de Pínos, funded by the Secretary of Universities and research (Government of Catalonia) and by the Horizon 2020 programme of research and innovation of the European union under the Marie Sktodowska-Curie grant agreement No 801370.
\clearpage

\begin{table}[th]
\centering
\smallskip 
 \resizebox{1.0\columnwidth}{!}{%
\begin{tabular}{lcccccccc}
 \hline
\toprule

Index & Mean & Median & SD & Skewness & Kurtosis & J-B & L-B & ADF \\ 
\hline
FTSE 100 & 0.086 & 0.234 & 2.396 & -1.456 & 14.717 & \textbf{17517.036} & \textbf{62.674} & \textbf{-19.729} \\ 
NIKKEI 225 & 0.046 & 0.237 & 2.946 & -0.748 & 6.421 & \textbf{3383.250} & \textbf{175.024} & \textbf{-19.159} \\ 
S\&P 500 & 0.163 & 0.320 & 2.338 & -0.947 & 7.367 & \textbf{4503.301} & \textbf{403.851} & \textbf{-20.325} \\   
\hline
{Correlation matrix}\\[0.05in]
& FTSE 100 & NIKKEI 225 & S\&P 500 \\ 
FTSE 100 & 1 & & \\    
NIKKEI 225 & 0.510 & 1 & \\     
S\&P 500 & 0.709 & 0.501 & 1 \\ 

\bottomrule 
\hline
\end{tabular}}
\caption{Summary statistics of weekly returns of the three indices for the entire sample from April 26, 1985 to February 01, 2021. The test statistics are displayed in boldface when the null hypothesis is rejected at the 1\% significance level. 
 J-B, L-B and ADF denote the Jarque-Bera test, the Ljung-Box test on squared returns with 4 lags and the Augmented Dickey-Fuller unit root test with 4 lags, respectively.}\label{tab:stat}
\end{table}

\begin{table}[!htbp]
\centering
 \smallskip 
 \resizebox{1.0\columnwidth}{!}{%
 \setlength\tabcolsep{3pt}
\begin{tabular}{lccccccccccccccc}
\toprule
$\boldsymbol{\tau}$ & \multicolumn{5}{c}{$[0.1, 0.1, 0.1]$} & \multicolumn{5}{c}{$[0.05, 0.05, 0.05]$} & \multicolumn{5}{c}{$[0.01, 0.01, 0.01]$} \\\cmidrule(r){2-6}\cmidrule(l){7-11}\cmidrule(l){12-16}
 & LR$_{uc}$ & LR$_{cc}$ & DQ & U$_{ES}$ & C$_{ES}$ & LR$_{uc}$ & LR$_{cc}$ & DQ & U$_{ES}$ & C$_{ES}$ & LR$_{uc}$ & LR$_{cc}$ & DQ & U$_{ES}$ & C$_{ES}$ \\
\hline
\\
\multicolumn{10}{l}{Panel A: multiplicative factor for the ES}\\ 
\multicolumn{10}{l}{SAV}\\ 
FTSE 100   & $\mathbf{0.043}$ & $\mathbf{0.455}$ & $\mathbf{7.194}$ & $\mathbf{1.467}$  & $\mathbf{7.735}$ & $\mathbf{2.658}$ & $\mathbf{3.469}$ & $\mathbf{8.288}$ & $\mathbf{0.128}$  & $\mathbf{6.711}$ & $\mathbf{0.756}$ & $\mathbf{1.111}$ & $\mathbf{6.350}$ & $5.063$           & $9.991$          \\
NIKKEI 225 & $5.571$          & $\mathbf{5.838}$ & $10.463$         & $\mathbf{0.022}$  & $\mathbf{6.553}$ & $\mathbf{1.851}$ & $\mathbf{2.806}$ & $\mathbf{7.047}$ & $\mathbf{0.543}$  & $11.086$         & $\mathbf{2.857}$ & $\mathbf{4.510}$ & $11.047$         & $\mathbf{-1.280}$ & $\mathbf{0.913}$ \\
S\&P 500   & $\mathbf{0.450}$ & $\mathbf{0.804}$ & $\mathbf{9.431}$ & $\mathbf{1.604}$  & $12.776$         & $\mathbf{1.203}$ & $\mathbf{3.890}$ & $\mathbf{8.917}$ & $\mathbf{0.931}$  & $12.748$         & $4.890$          & $\mathbf{5.815}$ & $30.388$         & $\mathbf{-1.229}$ & $9.568$          \\
\multicolumn{10}{l}{AS}\\ 
FTSE 100  & $\mathbf{1.067}$ & $\mathbf{1.210}$ & $\mathbf{6.641}$ & $\mathbf{0.663}$  & $\mathbf{8.773}$ & $\mathbf{1.203}$ & $\mathbf{2.314}$ & $\mathbf{6.266}$ & $\mathbf{0.434}$  & $\mathbf{5.243}$ & $\mathbf{1.096}$ & $\mathbf{4.184}$ & $\mathbf{3.351}$ & $\mathbf{-1.562}$ & $\mathbf{3.428}$ \\
NIKKEI 225& $\mathbf{3.571}$ & $\mathbf{4.932}$ & $\mathbf{1.064}$ & $\mathbf{0.261}$  & $\mathbf{8.043}$ & $\mathbf{2.658}$ & $\mathbf{3.469}$ & $\mathbf{8.094}$ & $\mathbf{0.579}$  & $\mathbf{9.193}$ & $\mathbf{1.747}$ & $\mathbf{2.019}$ & $\mathbf{8.332}$ & $\mathbf{-1.490}$ & $\mathbf{2.241}$ \\
S\&P 500  & $\mathbf{0.724}$ & $\mathbf{0.961}$ & $\mathbf{7.585}$ & $\mathbf{1.428}$  & $\mathbf{8.660}$ & $\mathbf{0.704}$ & $\mathbf{2.942}$ & $\mathbf{7.101}$ & $\mathbf{1.091}$  & $\mathbf{8.038}$ & $\mathbf{0.414}$ & $\mathbf{2.022}$ & $\mathbf{6.519}$ & $\mathbf{1.439}$  & $\mathbf{0.013}$ \\
\multicolumn{10}{l}{IG}\\
FTSE 100  & $\mathbf{0.450}$ & $\mathbf{0.456}$ & $\mathbf{5.705}$ & $\mathbf{1.011}$  & $9.926$          & $\mathbf{2.658}$ & $\mathbf{3.469}$ & $\mathbf{8.381}$ & $\mathbf{0.263}$  & $\mathbf{6.421}$ & $4.261$          & $\mathbf{4.399}$ & $9.853$          & $\mathbf{1.670}$  & $\mathbf{6.078}$ \\
NIKKEI 225& $6.547$          & $6.717$          & $11.814$         & $\mathbf{-0.295}$ & $\mathbf{8.787}$ & $\mathbf{1.851}$ & $\mathbf{2.806}$ & $\mathbf{6.954}$ & $\mathbf{0.597}$  & $\mathbf{9.478}$ & $\mathbf{1.747}$ & $\mathbf{2.019}$ & $\mathbf{5.738}$ & $\mathbf{-1.195}$ & $\mathbf{6.915}$ \\
S\&P 500  & $\mathbf{0.724}$ & $\mathbf{0.961}$ & $10.492$         & $\mathbf{1.298}$  & $12.164$         & $\mathbf{1.851}$ & $\mathbf{5.047}$ & $11.778$         & $\mathbf{0.579}$  & $\mathbf{7.730}$ & $\mathbf{0.952}$ & $\mathbf{1.309}$ & $\mathbf{6.009}$ & $\mathbf{1.699}$  & $\mathbf{3.150}$ \\
\hline
\\
\multicolumn{10}{l}{Panel B: AR formulation for the ES}\\ 
\multicolumn{10}{l}{SAV}\\ 
FTSE 100  & $4.589$          & $\mathbf{4.714}$ & $9.787$          & $\mathbf{0.184}$  & $\mathbf{4.671}$ & $\mathbf{2.587}$ & $\mathbf{3.490}$ & $11.336$         & $\mathbf{-1.699}$ & $\mathbf{2.256}$ & $\mathbf{1.055}$ & $\mathbf{2.922}$ & $\mathbf{9.309}$ & $-3.699$          & $\mathbf{2.256}$ \\
NIKKEI 225& $5.329$          & $\mathbf{5.431}$ & $10.613$         & $\mathbf{-1.570}$ & $\mathbf{7.718}$ & $\mathbf{1.547}$ & $\mathbf{2.297}$ & $\mathbf{7.295}$ & $\mathbf{-1.330}$ & $\mathbf{2.575}$ & $\mathbf{2.070}$ & $\mathbf{4.164}$ & $12.106$         & $4.330$           & $11.575$         \\
S\&P 500  & $6.840$          & $8.057$          & $12.670$         & $\mathbf{0.152}$  & $12.066$         & $\mathbf{1.476}$ & $\mathbf{3.516}$ & $\mathbf{2.591}$ & $1.969$           & $\mathbf{0.315}$ & $4.020$          & $\mathbf{5.144}$ & $15.297$         & $5.869$           & $\mathbf{0.315}$ \\
\multicolumn{10}{l}{AS}\\ 
FTSE 100  & $\mathbf{3.441}$ & $\mathbf{3.337}$ & $\mathbf{5.771}$ & $\mathbf{-1.594}$ & $\mathbf{1.128}$ & $\mathbf{2.249}$ & $\mathbf{5.271}$ & $\mathbf{3.573}$ & $\mathbf{1.470}$  & $\mathbf{7.960}$ & $\mathbf{3.182}$ & $\mathbf{2.913}$ & $\mathbf{3.248}$ & $\mathbf{-1.470}$ & $\mathbf{7.960}$ \\
NIKKEI 225& $\mathbf{3.651}$ & $\mathbf{5.168}$ & $\mathbf{6.650}$ & $\mathbf{-1.646}$ & $\mathbf{1.631}$ & $\mathbf{2.251}$ & $\mathbf{4.451}$ & $\mathbf{7.473}$ & $\mathbf{-1.350}$ & $\mathbf{2.542}$ & $\mathbf{0.177}$ & $\mathbf{4.879}$ & $\mathbf{3.313}$ & $\mathbf{1.350}$  & $\mathbf{2.542}$ \\
S\&P 500  & $\mathbf{2.250}$ & $\mathbf{3.371}$ & $\mathbf{1.745}$ & $\mathbf{0.523}$  & $\mathbf{0.008}$ & $\mathbf{1.535}$ & $\mathbf{3.321}$ & $\mathbf{6.457}$ & $\mathbf{1.103}$  & $\mathbf{6.304}$ & $\mathbf{2.922}$ & $\mathbf{3.042}$ & $\mathbf{6.036}$ & $\mathbf{-1.103}$ & $\mathbf{6.304}$ \\
\multicolumn{10}{l}{IG}\\
FTSE 100  & $5.214$          & $\mathbf{1.282}$ & $\mathbf{9.456}$ & $\mathbf{0.251}$  & $\mathbf{7.350}$ & $\mathbf{2.284}$ & $\mathbf{3.146}$ & $\mathbf{8.951}$ & $\mathbf{1.553}$  & $\mathbf{5.164}$ & $\mathbf{3.036}$ & $\mathbf{3.935}$ & $\mathbf{9.082}$ & $\mathbf{1.840}$  & $\mathbf{4.991}$ \\
NIKKEI 225& $12.095$         & $6.845$          & $\mathbf{6.904}$ & $\mathbf{1.285}$  & $9.997$          & $\mathbf{1.228}$ & $\mathbf{3.180}$ & $\mathbf{6.852}$ & $\mathbf{1.151}$  & $\mathbf{4.656}$ & $\mathbf{2.135}$ & $\mathbf{3.106}$ & $\mathbf{6.277}$ & $\mathbf{1.704}$  & $\mathbf{2.610}$ \\
S\&P 500  & $9.859$          & $\mathbf{1.564}$ & $\mathbf{6.456}$ & $\mathbf{1.732}$  & $\mathbf{4.639}$ & $\mathbf{1.731}$ & $\mathbf{4.584}$ & $10.327$         & $\mathbf{-1.703}$ & $\mathbf{1.925}$ & $\mathbf{1.041}$ & $\mathbf{5.188}$ & $15.358$         & $\mathbf{-1.898}$ & $\mathbf{0.013}$ \\
\bottomrule
\end{tabular}}
\caption{Marginal out-of-sample VaR and ES forecasts evaluation using the joint approach with the multiplicative factor in \eqref{deltaES} (Panel A) and the AR formulation in \eqref{ARES1}-\eqref{ARES2} (Panel B) for the ES. \textcolor{black}{At the 5\% significance level, } the critical values of the \textcolor{black}{LR$_{uc}$ and LR$_{cc}$ are $3.84$ and $5.99$, respectively. The U$_{ES}$ is rejected if the test statistic is greater (in absolute value) than $1.96$. Finally, the DQ test uses lagged violations at lag 4 while the C$_{ES}$ test considers the first 4 lagged autocorrelations, and the critical value for both is $9.49$.} The test statistics are displayed in boldface when the null hypotheses are not rejected at the 5\% significance level. 
}
\label{tab:back}
\end{table}

\begin{table}[!htbp]
\centering
 \smallskip 
 \resizebox{0.7\columnwidth}{!}{%
\begin{tabular}{lcccccc}
\toprule
$\boldsymbol{\tau}$ & \multicolumn{2}{c}{$[0.1, 0.1, 0.1]$} & \multicolumn{2}{c}{$[0.05, 0.05, 0.05]$} & \multicolumn{2}{c}{$[0.01, 0.01, 0.01]$} \\\cmidrule(r){2-3}\cmidrule(l){4-5}\cmidrule(l){6-7}
 & $S_\textnormal{FZ0}$ & $S_\textnormal{FZN}$ & $S_\textnormal{FZ0}$ & $S_\textnormal{FZN}$ & $S_\textnormal{FZ0}$ & $S_\textnormal{FZN}$ \\
\hline
\\
\multicolumn{7}{l}{Panel A: multiplicative factor for the ES}\\ 
\multicolumn{7}{l}{SAV}\\
FTSE 100   & $1.398$ & $2.661$ & $1.795$ & $3.221$ & $2.109$ & $3.399$ \\
NIKKEI 225 & $1.545$ & $3.091$ & $2.051$ & $3.666$ & $2.231$ & $3.678$ \\
S\&P 500   & $1.368$ & $2.552$ & $1.753$ & $3.126$ & $1.907$ & $3.215$ \\
\multicolumn{7}{l}{AS}\\ 
FTSE 100  & $1.371$ & $2.578$ & $1.750$ & $2.833$ & $1.954$ & $3.254$ \\
NIKKEI 225& $1.585$ & $2.890$ & $1.949$ & $3.155$ & $2.111$ & $3.636$ \\
S\&P 500  & $1.346$ & $2.431$ & $1.706$ & $2.690$ & $1.867$ & $3.142$ \\
\multicolumn{7}{l}{IG}\\ 
FTSE 100  & $1.383$ & $2.636$ & $1.765$ & $3.000$ & $1.946$ & $3.320$ \\
NIKKEI 225& $1.539$ & $3.025$ & $2.023$ & $3.387$ & $2.196$ & $3.689$ \\
S\&P 500  & $1.357$ & $2.535$ & $1.715$ & $2.943$ & $1.897$ & $3.257$ \\
\hline
\\
\multicolumn{7}{l}{Panel B: AR formulation for the ES}\\ 
\multicolumn{7}{l}{SAV}\\ 
FTSE 100  & $1.414$ & $2.878$ & $2.116$ & $2.362$ & $4.866$ & $5.740$ \\
NIKKEI 225& $1.766$ & $2.082$ & $2.241$ & $3.055$ & $5.220$ & $4.779$ \\
S\&P 500  & $1.406$ & $2.767$ & $1.940$ & $3.177$ & $4.929$ & $5.241$ \\
\multicolumn{7}{l}{AS}\\ 
FTSE 100  & $1.496$ & $1.845$ & $1.985$ & $2.325$ & $6.998$ & $5.009$ \\
NIKKEI 225& $1.714$ & $2.059$ & $2.315$ & $3.731$ & $7.233$ & $4.120$ \\
S\&P 500  & $1.478$ & $2.756$ & $2.048$ & $2.944$ & $6.929$ & $4.934$ \\
\multicolumn{7}{l}{IG}\\ 
FTSE 100  & $1.411$ & $1.849$ & $2.112$ & $2.800$ & $3.201$ & $5.025$ \\
NIKKEI 225& $1.671$ & $1.993$ & $2.305$ & $3.076$ & $3.432$ & $5.629$ \\
S\&P 500  & $1.370$ & $2.764$ & $2.031$ & $3.007$ & $3.164$ & $4.983$ \\
\bottomrule
\end{tabular}}
\caption{\textcolor{black}{Marginal out-of-sample VaR and ES forecasts evaluation based on average losses using the scoring functions in \eqref{lossfzn} and \eqref{lossfzg} for the joint approach with the multiplicative factor in \eqref{deltaES} (Panel A) and the AR formulation in \eqref{ARES1}-\eqref{ARES2} (Panel B) for the ES.}}
\label{tab:back_score}
\end{table}

\newgeometry{left=05mm,right=05mm}
\begin{table}[h]
\centering
\smallskip 
\resizebox{0.9\columnwidth}{!}{%
\begin{tabular}{lcccccccccc}
\hline
\toprule
$\bs{\tau}$ & \multicolumn{3}{c}{$[0.1, 0.1, 0.1]$} & \multicolumn{3}{c}{$[0.05, 0.05, 0.05]$} & \multicolumn{3}{c}{$[0.01, 0.01, 0.01]$} \\\cmidrule(r){2-4}\cmidrule(l){5-7}\cmidrule(l){8-10}
& CAViaR-SAV & CAViaR-AS & CAViaR-IG & CAViaR-SAV & CAViaR-AS & CAViaR-IG & CAViaR-SAV & CAViaR-AS & CAViaR-IG \\ 
\hline \\[-0.05in]
\multicolumn{6}{l}{Panel A: multiplicative factor for the ES} \\[0.05in]
CAViaR-SAV & - & 3.169 & 2.020 & - & 4.090 & 4.630 & - & 3.832 & 3.777 \\ 
           & - & (0.999) & (0.978) & - & (1.000) & (1.000) & - & (1.000) & (1.000) \\ 
CAViaR-AS  & -3.169 & - & -2.575 & -4.090 & - & -2.394 & -3.832 & - & -1.931 \\  
           & (0.001) & - & (0.005) & (0.000) & - & (0.009) & (0.000) & - & (0.027) \\ 
CAViaR-IG  & -2.020 & 2.575 & - & -4.630 & 2.394 & - & -3.777 & 1.931 & - \\ 
           & (0.022) & (0.995) & - & (0.000) & (0.991) & - & (0.000) & (0.973) & - \\ 
\hline \\
\multicolumn{6}{l}{Panel B: AR formulation for the ES} \\[0.05in]
CAViaR-SAV & - & 2.635 & 1.919 & - & 3.688 & 4.114 & - & 3.927 & 4.728 \\ 
           & - & (0.996) & (0.972) & - & (1.000) & (1.000) & - & (1.000) & (1.000) \\ 
CAViaR-AS  & -2.635 & - & -2.320 & -3.688 & - & 6.090 & -3.927 & - & 6.273 \\  
           & (0.004) & - & (0.010) & (0.000) & - & (1.000) & (0.000) & - & (1.000) \\ 
CAViaR-IG  & -1.919 & 2.320 & - & -4.114 & -6.090 & - & -4.728 & -6.273 & - \\ 
           & (0.028) & (0.990) & - & (0.000) & (0.000) & - & (0.000) & (0.000) & - \\ 
\bottomrule
\hline
\end{tabular}}
\caption{Test statistics and $p$-values (in parentheses) of the \cite{diebold2002comparing} pairwise test between competing CAViaR models in predicting one-week-ahead returns using the joint approach with the multiplicative factor in \eqref{deltaES} (Panel A) and the AR formulation in \eqref{ARES1}-\eqref{ARES2} (Panel B) for the ES. In each panel, the null hypothesis is that, on average, the forecasts obtained with model $i$ are not statistically different from the ones obtained with model $j$, using the multivariate scoring rule $S_{MAL_{t}}^{(j)}(\bs{\tau})$ in \eqref{lossmal}.}
\label{tab:dieboldMAL}
\end{table}
\restoregeometry

\begin{table}[!t]
 \centering
 \smallskip 
\begin{tabular}{lccc}
\toprule
$\bs{\tau}$ & \multicolumn{1}{c}{$[0.1, 0.1, 0.1]$} & \multicolumn{1}{c}{$[0.05, 0.05, 0.05]$} & \multicolumn{1}{c}{$[0.01, 0.01, 0.01]$}  \\\cmidrule(r){2-2}\cmidrule(l){3-3}\cmidrule(l){4-4}
& \multicolumn{3}{c}{\textit{AR formulation}} \\\cmidrule(r){2-4}
& CAViaR-AS & CAViaR-AS & CAViaR-AS \\
\hline \\[-0.05in]
\multicolumn{4}{l}{\textit{Multiplicative factor}} \\\cmidrule(r){1-2} \\
CAViaR-AS  &-1.997 & -6.273 & -5.029 \\
           & (0.023) & (0.000) & (0.000) \\
\bottomrule
\end{tabular}
   \caption{Test statistics and $p$-values (in parentheses) of the \cite{diebold2002comparing} pairwise test between the CAViaR-AS specifications using the joint approach with the constant multiplicative factor in \eqref{deltaES} and the AR formulation in \eqref{ARES1}-\eqref{ARES2} for the ES component in predicting one-week-ahead returns. The null hypothesis is that the two approaches have the same forecasting performance.}
   \label{tab:diebold2}
\end{table}

\newgeometry{left=05mm,right=05mm}
\begin{table}[h]
\centering
\smallskip 
\resizebox{0.9\columnwidth}{!}{%
\begin{tabular}{lcccccccccc}
\hline
\toprule
$\bs{\tau}$ & \multicolumn{3}{c}{$[0.1, 0.1, 0.1]$} & \multicolumn{3}{c}{$[0.05, 0.05, 0.05]$} & \multicolumn{3}{c}{$[0.01, 0.01, 0.01]$} \\\cmidrule(r){2-4}\cmidrule(l){5-7}\cmidrule(l){8-10}           
& CAViaR-SAV & CAViaR-AS & CAViaR-IG & CAViaR-SAV & CAViaR-AS & CAViaR-IG & CAViaR-SAV & CAViaR-AS & CAViaR-IG \\ 
\hline \\[-0.05in]
\multicolumn{6}{l}{Panel A: multiplicative factor for the ES} \\[0.05in]
CAViaR-SAV & - & 1.383 & 0.971 & - & 1.905 & 0.938 & - & 0.920 & 0.274 \\  
           & - & (0.916) & (0.834) & - & (0.971) & (0.826) & - & (0.821) & (0.608) \\  
CAViaR-AS  & -1.383 & - & -1.030 & -1.905 & - & -1.936 & -0.920 & - & -0.975 \\  
           & (0.084) & - & (0.152) & (0.029) & - & (0.027) & (0.179) & - & (0.165) \\  
CAViaR-IG  & -0.971 & 1.030 & - & -0.938 & 1.936 & - & -0.274 & 0.975 & - \\  
           & (0.166) & (0.848) & - & (0.174) & (0.973) & - & (0.392) & (0.835) & - \\  
\hline \\
\multicolumn{6}{l}{Panel B: AR formulation for the ES} \\[0.05in]
CAViaR-SAV & - & 0.179 & 0.273 & - & 0.473 & 0.111 & - & 0.188 & -0.740 \\  
           & - & (0.571) & (0.608) & - & (0.682) & (0.544) & - & (0.575) & (0.230) \\  
CAViaR-AS  & -0.179 & - & -0.003 & -0.473 & - & -0.537 & -0.188 & - & -0.520 \\  
           & (0.429) & - & (0.499) & (0.318) & - & (0.296) & (0.425) & - & (0.302) \\  
CAViaR-IG  & -0.273 & 0.003 & - & -0.111 & 0.537 & - & 0.740 & 0.520 & - \\  
           & (0.392) & (0.501) & - & (0.456) & (0.704) & - & (0.770) & (0.698) & - \\  
\bottomrule
\hline
\end{tabular}}
\caption{Test statistics and $p$-values (in parentheses) of the \cite{diebold2002comparing} pairwise test between competing CAViaR models in predicting one-week-ahead returns using the univariate approach of \cite{taylor2017forecasting} with the multiplicative factor in \eqref{deltaES} (Panel A) and the AR formulation in \eqref{ARES1}-\eqref{ARES2} (Panel B) for the ES. In each panel, the null hypothesis is that, on average, the forecasts obtained with model $i$ are not statistically different from the ones obtained with model $j$, using the $S_{AL_{t}}^{(j)}(\bs{\tau})$ scoring rule in \eqref{lossAL}.}
\label{tab:dieboldAL}
\end{table}
\restoregeometry

\clearpage
\newgeometry{left=05mm,right=05mm}
\begin{table}[!h]
 \centering
 \smallskip 
 \resizebox{1.0\columnwidth}{!}{%

\begin{tabular}{lccccccccc}
\hline
\toprule

$\bs{\tau}$ & \multicolumn{3}{c}{$[0.1, 0.1, 0.1]$} & \multicolumn{3}{c}{$[0.05, 0.05, 0.05]$} & \multicolumn{3}{c}{$[0.01, 0.01, 0.01]$}  \\\cmidrule(r){2-4}\cmidrule(l){5-7}\cmidrule(l){8-10}

& \multicolumn{9}{c}{\textit{Univariate approach}} \\\cmidrule(r){2-10}

 & CAViaR-SAV & CAViaR-AS & CAViaR-IG & CAViaR-SAV & CAViaR-AS & CAViaR-IG & CAViaR-SAV & CAViaR-AS & CAViaR-IG \\ 
\hline \\[-0.05in]
\multicolumn{1}{l}{\textit{Joint approach}} \\\cmidrule(r){1-1} \\
\multicolumn{10}{l}{Panel A: multiplicative factor for the ES} \\[0.05in]
 CAViaR-SAV & -4.368 & -4.432 & -4.488 & -2.966 & -2.748 & -2.890 & -3.166 & -3.348 & -3.290 \\ 
  & (0.000) & (0.000) & (0.000) & (0.002) & (0.003) & (0.002) & (0.002) & (0.003) & (0.002)\\  
  CAViaR-AS & -4.386 & -4.561 & -4.601 & -3.108 & -2.973 & -3.127 & -3.066 & -2.748 & -3.190 \\ 
  & (0.000) & (0.000) & (0.000) & (0.001) & (0.001) & (0.001) & (0.001) & (0.003) & (0.001) \\  
  CAViaR-IG & -4.389 & -4.523 & -4.573 & -3.063 & -2.880 & -3.029 & -2.656 & -2.048 & -2.690 \\ 
 & (0.000) & (0.000) & (0.000) & (0.001) & (0.002) & (0.001) & (0.004) & (0.020) & (0.005) \\  
 \hline \\
 \multicolumn{6}{l}{Panel B: AR formulation for the ES} \\[0.05in]
  CAViaR-SAV & -8.913 & -8.622 & -8.764 & -6.290 & -6.104 & -6.118 & -4.467 & -4.859 & -4.433 \\ 
  & (0.000) & (0.000) & (0.000) & (0.000) & (0.000) & (0.000) & (0.000) & (0.000) & (0.000) \\  
  CAViaR-AS & -8.536 & -8.278 & -8.417 & -6.413 & -6.472 & -6.306 & -4.856 & -5.353 & -4.856 \\ 
  & (0.000) & (0.000) & (0.000) & (0.000) & (0.000) & (0.000) & (0.000) & (0.000) & (0.000) \\  
  CAViaR-IG & -8.737 & -8.440 & -8.613 & -6.530 & -6.467 & -6.394 & -4.894 & -5.336 & -4.852 \\ 
 & (0.000) & (0.000) & (0.000) & (0.000) & (0.000) & (0.000) & (0.000) & (0.000) & (0.000) \\  
  \bottomrule
 \hline
   \end{tabular}}
   \caption{Test statistics and $p$-values (in parentheses) of the \cite{diebold2002comparing} pairwise test between the competing joint and univariate approaches in predicting one-week-ahead returns with the multiplicative factor in \eqref{deltaES} (Panel A) and the AR formulation in \eqref{ARES1}-\eqref{ARES2} (Panel B) for the ES. The null hypothesis is that the two approaches have the same forecasting performance.}
   \label{tab:dieboldmarg}
\end{table}
\restoregeometry

\begin{table}[htbp]
\centering
 \smallskip 

 \resizebox{0.7\columnwidth}{!}{%
\begin{tabular}{lccccccc}
\hline
\toprule
Portfolio & Mean & SD & \textcolor{black}{$S_\textnormal{FZ0}$} & \textcolor{black}{$S_\textnormal{FZN}$} & $S_\textnormal{AL}$ & \textcolor{black}{SR} & \textcolor{black}{HHI}\\
\hline
\multicolumn{8}{l}{Panel A: $\bs{\tau} = [0.1, 0.1, 0.1]$}\\[0.05in]
SMV & -2.716 & 0.883 & 1.317 & 1.549 & 2.473 & 0.225 & 0.558 \\
Mom-SMV & -2.785 & 0.859 & 1.405 & 1.592 & 2.486 & 0.221 & 0.563 \\
MV-G-DCC-N & -2.519 & 1.175 & 1.399 & 1.621 & 2.516 & 0.192 & 0.680 \\
MV-G-DCC-t & -2.391 & 1.107 & 1.380 & 1.911 & 2.491 & 0.198 & 0.662 \\
MV-G-aDCC-N & -2.520 & 1.174 & 1.399 & 1.622 & 2.491 & 0.195 & 0.697 \\
MV-G-aDCC-t & -2.392 & 1.107 & 1.379 & 1.912 & 2.487 & 0.193 & 0.673 \\
MV-G-DCC-SN & -3.198 & 1.483 & 1.419 & 1.876 & 2.511 & 0.187 & 0.693 \\
MV-G-DCC-St & -3.582 & 1.674 & 1.448 & 2.216 & 2.544 & 0.197 & 0.676 \\
MV-G-aDCC-SN & -3.199 & 1.482 & 1.418 & 1.876 & 2.511 & 0.188 & 0.691 \\
MV-G-aDCC-St & -3.582 & 1.633 & 1.459 & 2.208 & 2.544 & 0.200 & 0.678 \\
\hline
\multicolumn{8}{l}{Panel B: $\bs{\tau} = [0.05, 0.05, 0.05]$}\\[0.05in]
SMV & -4.016 & 1.240 & 1.563 & 1.612 & 2.529 & 0.204 & 0.557 \\
Mom-SMV & -3.759 & 1.244 & 1.592 & 1.615 & 2.615 & 0.189 & 0.562 \\
MV-G-DCC-N & -3.232 & 1.508 & 1.696 & 1.624 & 2.735 & 0.192 & 0.680 \\
MV-G-DCC-t & -3.174 & 1.476 & 1.655 & 2.024 & 2.696 & 0.198 & 0.662 \\
MV-G-aDCC-N & -3.234 & 1.507 & 1.695 & 1.625 & 2.735 & 0.195 & 0.697 \\
MV-G-aDCC-t & -3.176 & 1.476 & 1.655 & 2.025 & 2.696 & 0.193 & 0.673 \\
MV-G-DCC-SN & -3.812 & 1.770 & 1.666 & 1.868 & 2.706 & 0.187 & 0.693 \\
MV-G-DCC-St & -4.429 & 2.082 & 1.668 & 2.336 & 2.711 & 0.197 & 0.676 \\
MV-G-aDCC-SN & -3.809 & 1.765 & 1.664 & 1.871 & 2.705 & 0.188 & 0.691 \\
MV-G-aDCC-St & -4.430 & 2.027 & 1.679 & 2.327 & 2.722 & 0.200 & 0.678 \\
\hline

\multicolumn{8}{l}{Panel C: $\bs{\tau} = [0.01, 0.01, 0.01]$}\\[0.05in]
SMV & -5.227 & 1.936 & 1.989 & 1.656 & 3.031 & 0.154 & 0.533\\
Mom-SMV & -5.346 & 2.036 & 2.018 & 1.756 & 3.043 & 0.141 & 0.527\\
MV-G-DCC-N & -4.570 & 2.132 & 2.523 & 2.201 & 3.524 & 0.192 & 0.680 \\
MV-G-DCC-t & -4.913 & 2.312 & 2.263 & 2.073 & 3.266 & 0.198 & 0.662 \\
MV-G-aDCC-N & -4.572 & 2.132 & 2.523 & 2.202 & 3.522 & 0.195 & 0.697 \\
MV-G-aDCC-t & -4.914 & 2.311 & 2.263 & 2.074 & 3.266 & 0.193 & 0.673 \\
MV-G-DCC-SN & -5.004 & 2.325 & 2.406 & 1.865 & 3.407 & 0.187 & 0.693 \\
MV-G-DCC-St & -6.405 & 3.054 & 2.125 & 2.534 & 3.129 & 0.197 & 0.676 \\
MV-G-aDCC-SN & -5.009 & 2.336 & 2.395 & 1.877 & 3.396 & 0.188 & 0.691 \\
MV-G-aDCC-St & -6.395 & 2.948 & 2.145 & 2.503 & 3.149 & 0.200 & 0.678 \\
\bottomrule
\hline
   \end{tabular}}
   \caption{\small Evaluation of portfolios VaR and ES out-of-sample forecasts. 
 Mean and SD report the average and standard deviation of portfolio VaR. $S_\textnormal{FZ0}$, $S_\textnormal{FZN}$ and $S_\textnormal{AL}$ show the average losses using the scoring functions of \cite{Patton2019}, \cite{nolde2017elicitability} and \cite{taylor2017forecasting} in \eqref{lossfzg}, \eqref{lossfzn} and \eqref{lossALu}, respectively. SR and HHI denote the portfolio Sharpe Ratio and the averaged Herfindahl-Hirschman Index.
}\label{tab:pfback}
\end{table}

\clearpage
\clearpage
\newgeometry{left=10mm,right=10mm}
\begin{figure}
\begin{subfigure}[a] {\textwidth} 
\includegraphics[width=1\linewidth, height=5cm, width=6.5cm]{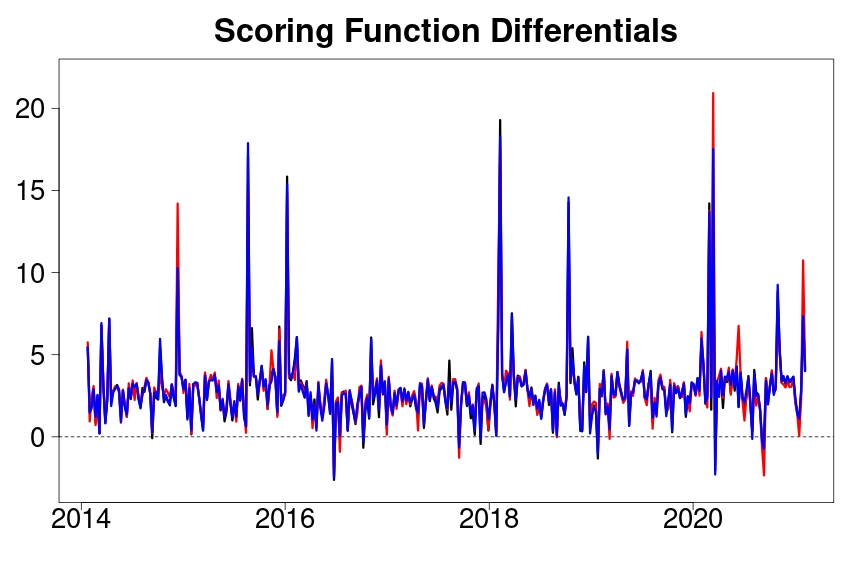}
\includegraphics[width=1\linewidth, height=5cm, width=6.5cm]{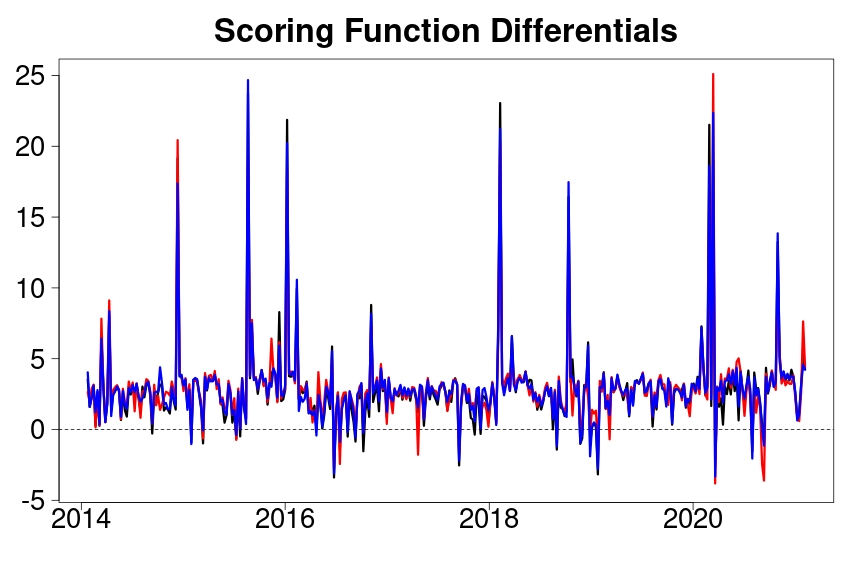}
\includegraphics[width=1\linewidth, height=5cm, width=6.5cm]{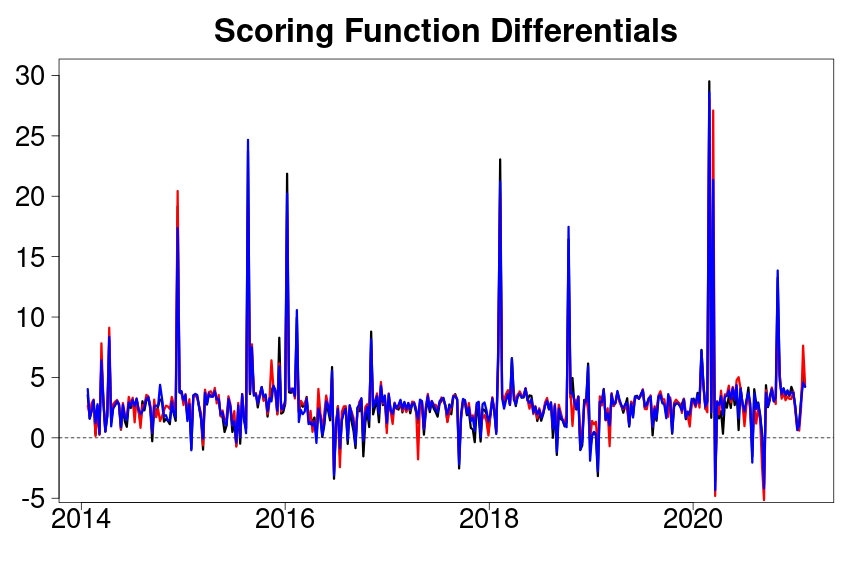}
\end{subfigure}
\caption{Scoring function differentials, $S_{AL_{t}}(\bs{\tau}) - S_{MAL_{t}}(\bs{\tau})$, between the $S_{AL_{t}}(\bs{\tau})$ loss in \eqref{lossAL} of the univariate approach of \cite{taylor2017forecasting} and the $S_{MAL_{t}}(\bs{\tau})$ loss in \eqref{lossmal} for the joint method, over the out-of-sample period at $\bs{\tau} = [0.1, 0.1, 0.1]$ (left plot), $\bs{\tau} = [0.05, 0.05, 0.05]$ (center plot) and $\bs{\tau} = [0.01, 0.01, 0.01]$ (right plot) for the CAViaR-SAV (black), CAViaR-AS (red) and CAViaR-IG (blue) specifications, with the ES modeled as in \eqref{deltaES}.}
\label{diff}
\end{figure}
\restoregeometry

\newgeometry{left=10mm,right=10mm}
\begin{figure}
\begin{subfigure}[a] {\textwidth} 
\includegraphics[width=1\linewidth, height=5cm, width=6.5cm]{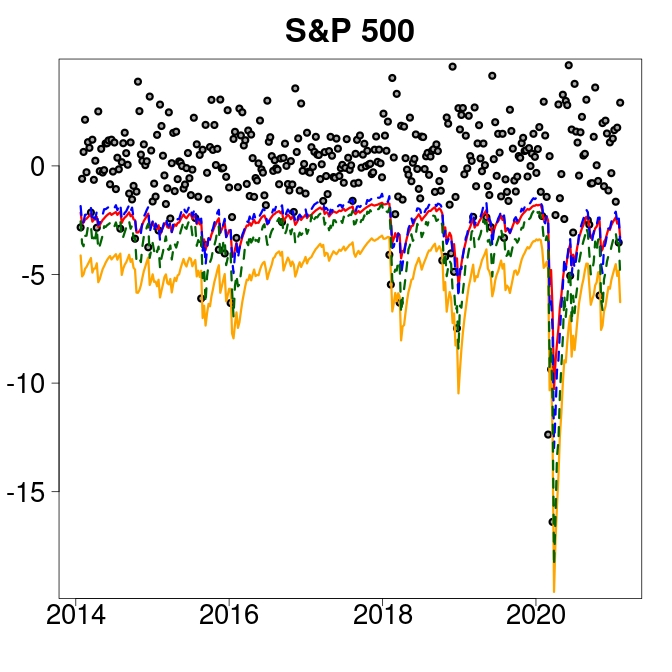}
\includegraphics[width=1\linewidth, height=5cm, width=6.5cm]{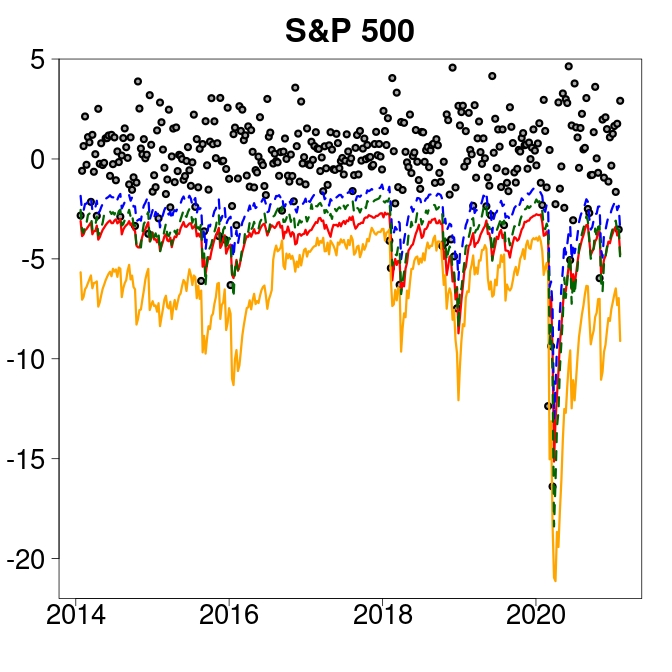}
\includegraphics[width=1\linewidth, height=5cm, width=6.5cm]{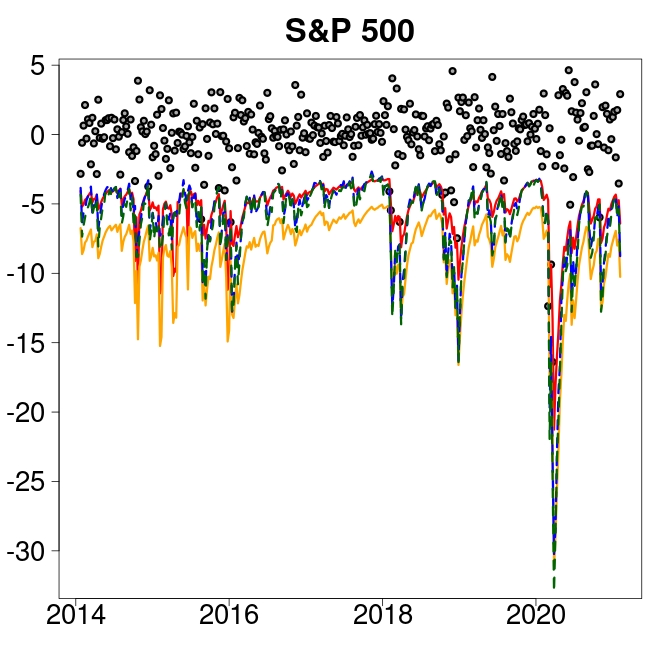}
\end{subfigure}
\begin{subfigure}[b] {\textwidth} 
\includegraphics[width=1\linewidth, height=5cm, width=6.5cm]{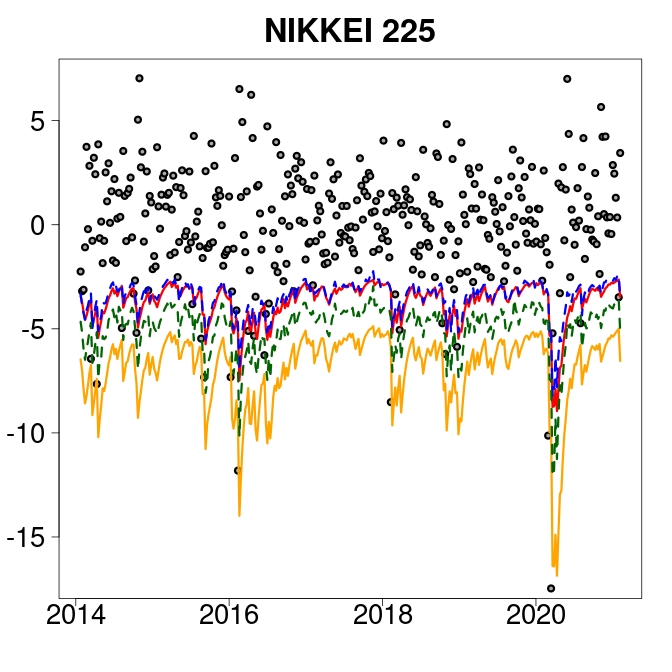}
\includegraphics[width=1\linewidth, height=5cm, width=6.5cm]{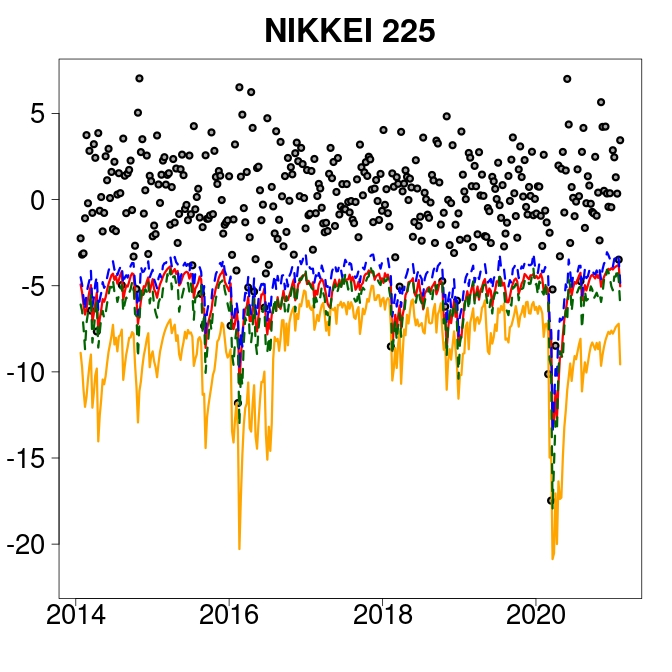}
\includegraphics[width=1\linewidth, height=5cm, width=6.5cm]{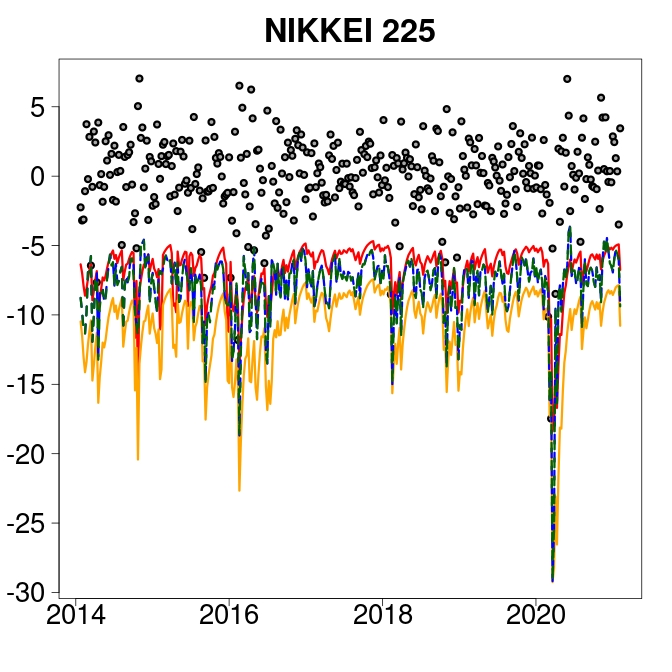}
\end{subfigure}
\begin{subfigure}[b] {\textwidth} 
\includegraphics[width=1\linewidth, height=5cm, width=6.5cm]{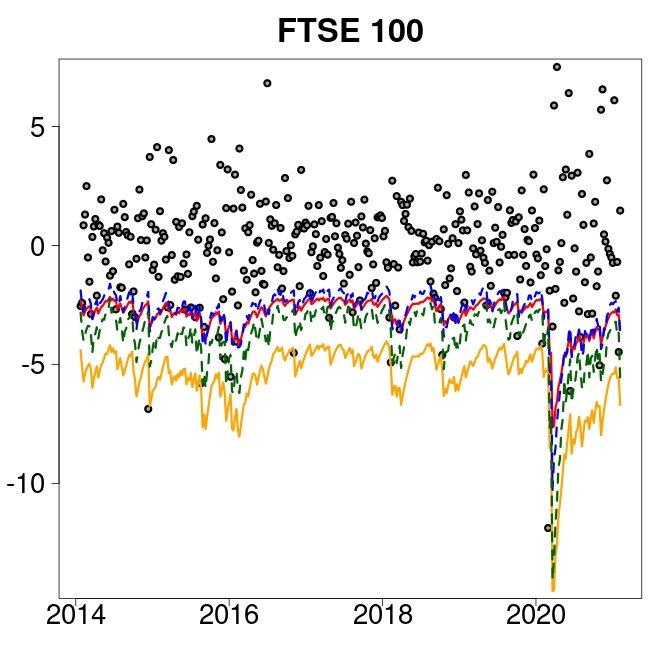}
\includegraphics[width=1\linewidth, height=5cm, width=6.5cm]{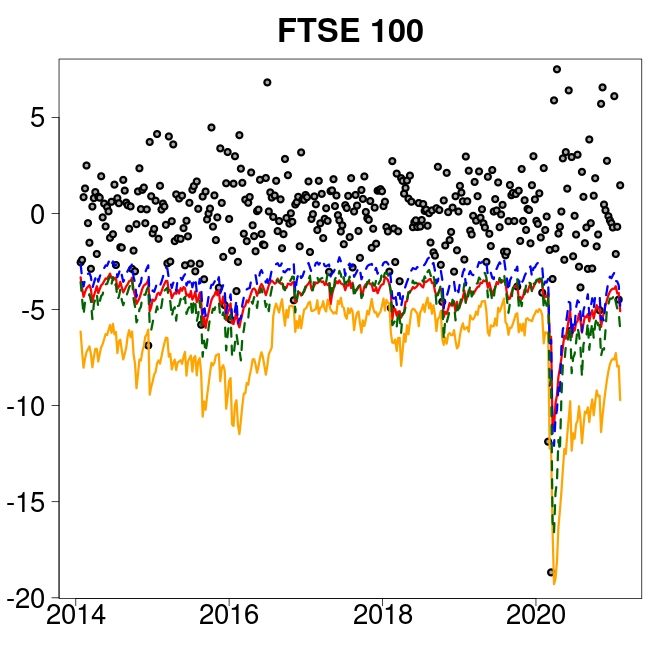}
\includegraphics[width=1\linewidth, height=5cm, width=6.5cm]{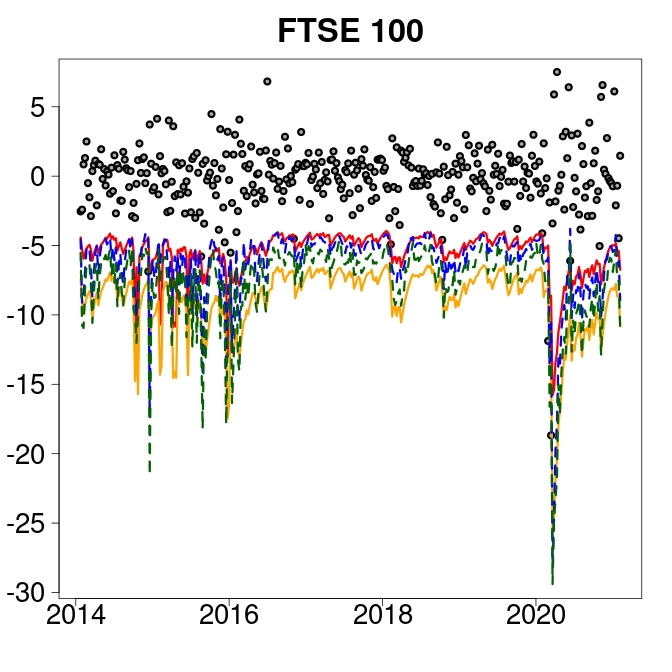}
\end{subfigure}
\caption{Out-of-sample forecasts of VaR and ES for the three stock indices, estimated with the CAViaR-AS specification using both the univariate and the joint approaches, and the ES modeled as in \eqref{deltaES}. The dotted blue and the solid red lines refer to the VaR predictions, estimated  with the univariate and the multiple approach, respectively. The estimated ES is represented by the dotted green line (for univariate method of \cite{taylor2017forecasting}) and the solid orange line (for the multivariate approach). Left panels refer to $\bs{\tau} = [0.1, 0.1, 0.1]$, the center panels refer to $\bs{\tau} = [0.05, 0.05, 0.05]$ while the case of $\bs{\tau} = [0.01, 0.01, 0.01]$ is displayed in the right panels. The gray dots represent the observed weekly returns for the considered stock index. }
\label{VARandES}
\end{figure}
\restoregeometry

\newgeometry{left=10mm,right=10mm}
\begin{figure}
\begin{subfigure}[a] {\textwidth} 
\includegraphics[width=1\linewidth, height=5cm, width=6.5cm]{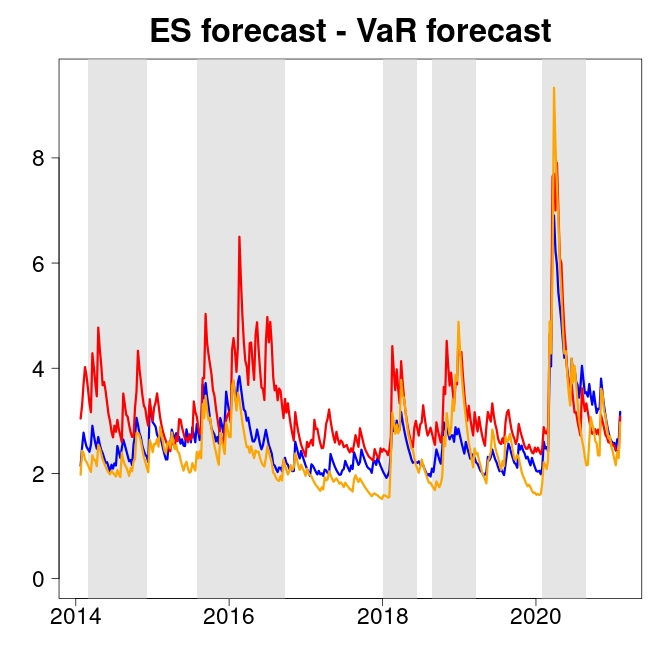}
\includegraphics[width=1\linewidth, height=5cm, width=6.5cm]{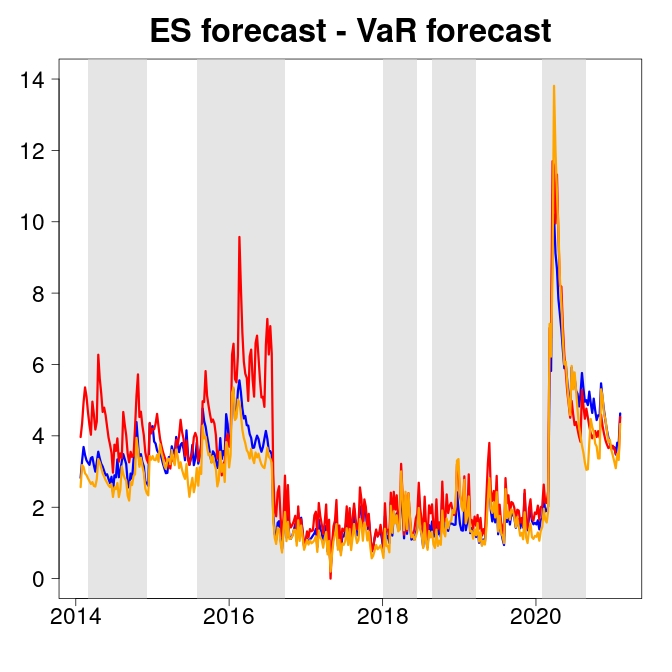}
\includegraphics[width=1\linewidth, height=5cm, width=6.5cm]{DIFF10_v3.jpeg}
\end{subfigure}
\begin{subfigure}[b] {\textwidth} 
\includegraphics[width=1\linewidth, height=5cm, width=6.5cm]{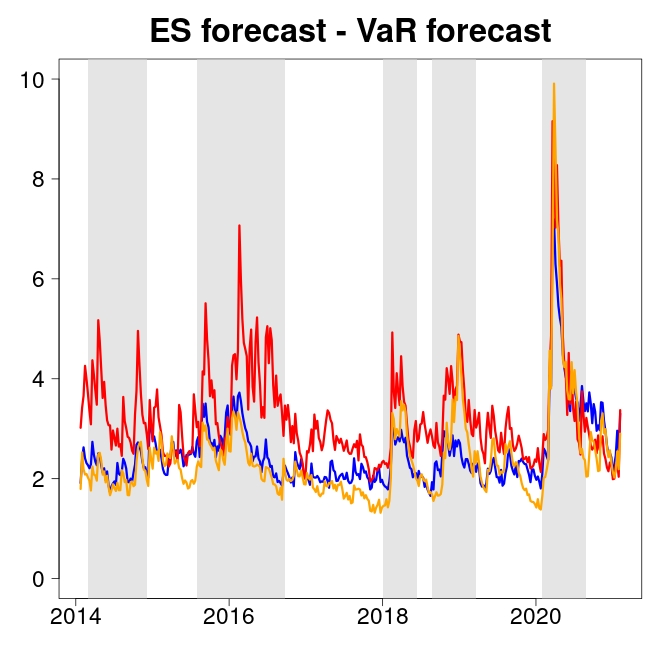}
\includegraphics[width=1\linewidth, height=5cm, width=6.5cm]{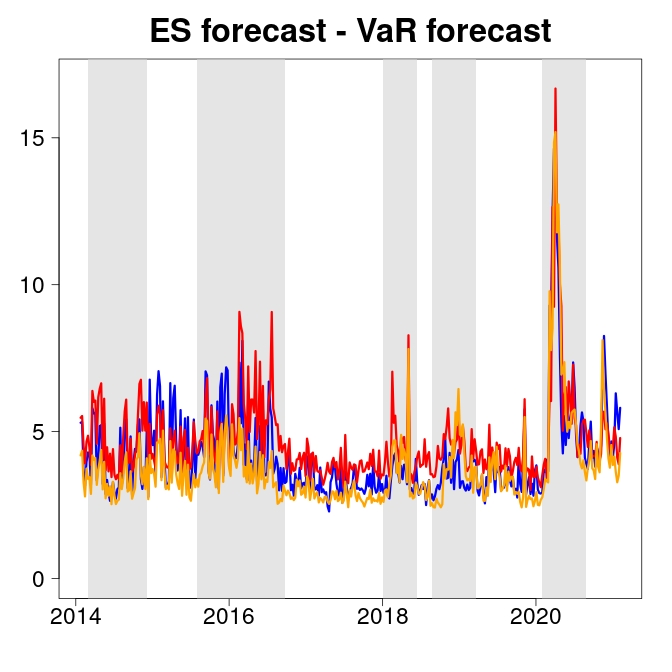}
\includegraphics[width=1\linewidth, height=5cm, width=6.5cm]{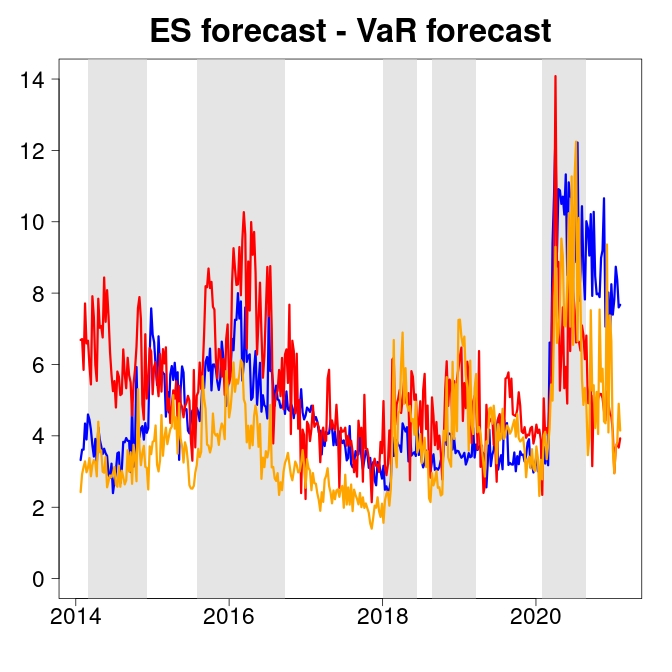}
\end{subfigure}
\caption{Absolute difference between the out-of-sample VaR and ES forecasts for the three stock indices, estimated with the CAViaR-AS specification, and using the ES modeled both as in \eqref{deltaES} (first row) and with the AR specification of \eqref{ARES1}-\eqref{ARES2} (second row), at the $\bs{\tau} = [0.1, 0.1, 0.1]$ (left column), $\bs{\tau} = [0.05, 0.05, 0.05]$ (center column) and $\bs{\tau} = [0.01, 0.01, 0.01]$ (right column) quantile levels. The blue, red and orange lines refer to the FTSE 100, NIKKEI 225 and S\&P 500 stock market indices, respectively. The grey bands correspond to the recession dates and to various economic and financial crises occurred in: 2014,03-2015,02; 2015,07-2016,09; 2018,01-2018,06; 2018,08-2019,03; 2020,02-2020,03.}
\label{VARandES2}
\end{figure}
\restoregeometry


\newgeometry{left=10mm,right=10mm}
\begin{figure}
\centering
\begin{subfigure}[a] {\textwidth} 
\includegraphics[width=1\linewidth, height=5.3cm, width=6.5cm]{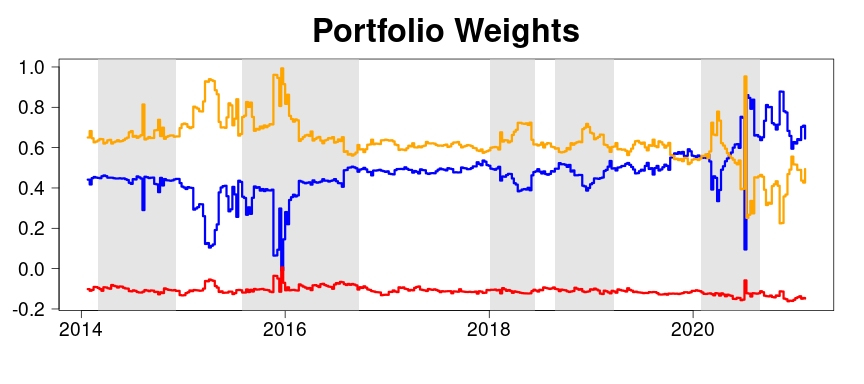}
\includegraphics[width=1\linewidth, height=5.3cm, width=6.5cm]{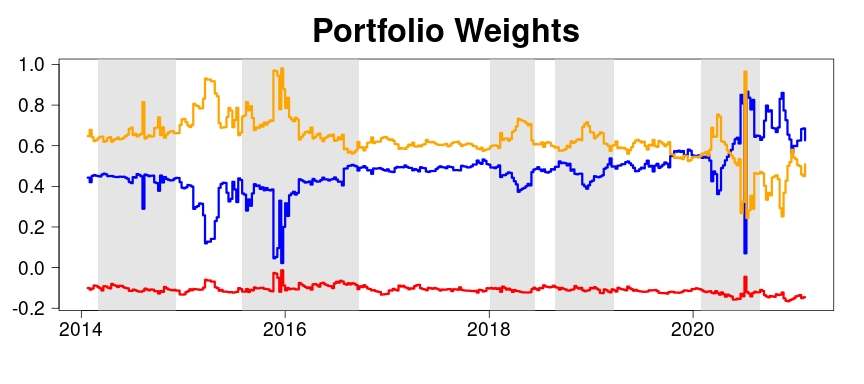}
\includegraphics[width=1\linewidth, height=5.3cm, width=6.5cm]{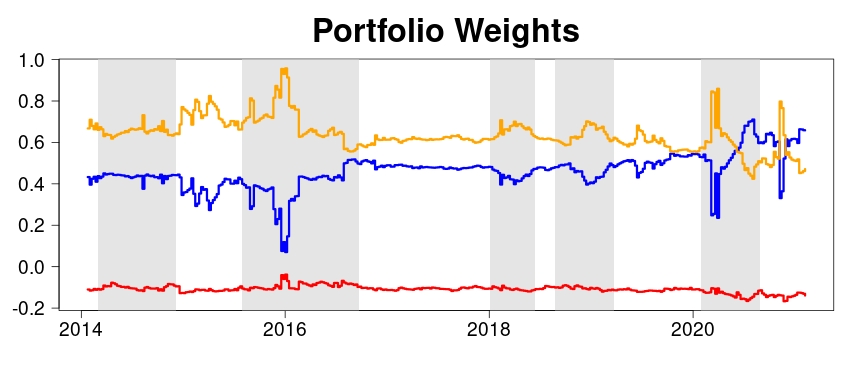}
\end{subfigure}
\caption{Optimal portfolio weights paths over the out-of-sample period computed using the selected CAViaR-AS model at $\tau = 0.1$ (left panel), $\tau = 0.05$ (central panel) and $\tau = 0.01$ (right panel). The optimal portfolios weights comprise the FTSE 100 (blue), NIKKEI 225 (red) and S\&P 500 (orange) stock market indices. The grey bands correspond to the recession dates and to various economic and financial crises occurred in: 2014,03-2015,02; 2015,07-2016,09; 2018,01-2018,06; 2018,08-2019,03; 2020,02-2020,03.}
\label{portfolioweights}
\end{figure}
\restoregeometry

\begin{figure}
\centering
\begin{subfigure}[a] {\textwidth} 
\centering
\includegraphics[width=1\linewidth, height=6.5cm, width=10cm]{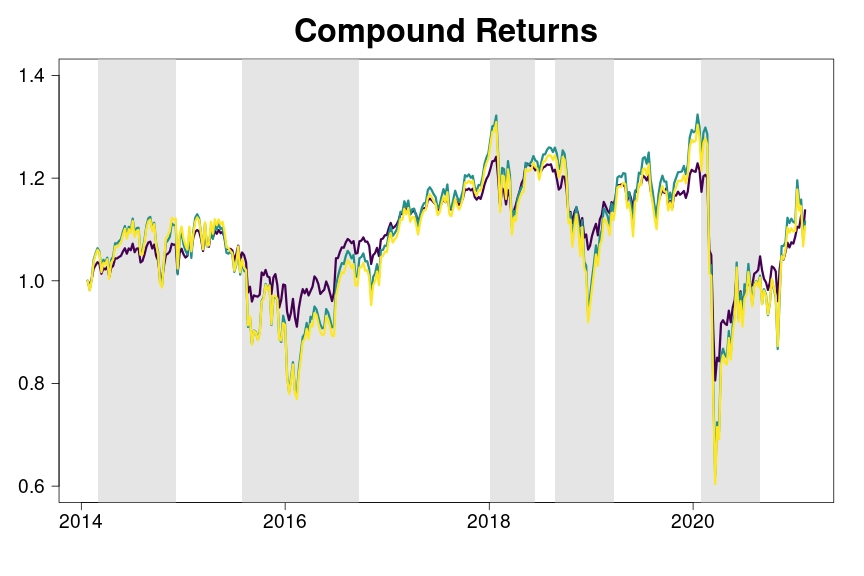}
\end{subfigure}
\caption{Compound returns over the out-of-sample period computed using the selected CAViaR-AS model at $\tau = 0.1$ (violet), $\tau = 0.05$ (green) and $\tau = 0.01$ (yellow). The grey bands correspond to the recession dates and to various economic and financial crises occurred in: 2014,03-2015,02; 2015,07-2016,09; 2018,01-2018,06; 2018,08-2019,03; 2020,02-2020,03.}
\label{compound}
\end{figure}

\clearpage
\section*{Appendix A. Proof of Proposition \ref{ALDuni} }



 As stated in \cite{petrella2018joint}, the $\mbox{MAL}_p(\bs \mu, \bs D \bs{\tilde  \xi}, \bs D \bs {\tilde\Sigma} \bs D)$ in \eqref{MALdensityConstr} can be written as a location-scale mixture, having the following representation:
\ber \label{mixtureALD2}
\bs Y=  \bs \mu   + \bs D \bs {\tilde \xi} W + \sqrt{W}  \bs D \bs {\tilde\Sigma}^{1/2}\bs Z
\eer  

\noindent {where $\bs {\tilde\xi}= [\tilde \xi_1, \tilde \xi_2,...,\tilde  \xi_p]'$, having generic element  $\tilde \xi_j= \frac{1- 2 \tau_j}{\tau_j(1 - \tau_j)}$, $j=1,..., p$. $\bs {\tilde \Sigma}$ is a $p \times p$ positive definite matrix such that $\bs {\tilde \Sigma} = \bs{\tilde \Lambda} \bs \Psi \bs{\tilde \Lambda}$, with $\bs \Psi $ being a correlation matrix and $\bs{\tilde \Lambda}= \diag[\tilde \sigma_1, \tilde \sigma_1,..., \tilde \sigma_p]$, with $\tilde \sigma_j^2= \frac{2}{\tau_j (1 - \tau_j)}$, $j=1,..., p$. Finally, $\bs Z \sim {\cal N}_p(\bs 0_p, \bs I_p)$ denotes a $p$-variate standard Normal distribution and  $W \sim \mbox{Exp}(1)$ has a standard Exponential distribution, with $\bs Z$ being independent of $W$. Notice that, under the constraints imposed on $\bs {\tilde\xi}$ and $\bs{\tilde \Lambda}$, the representation in \eqref{mixtureALD2} implies that:}
\ber \label{mixturecondALD2}
\bs Y \mid W = w \sim {\cal N}_p(\bs \mu + \bs D \bs {\tilde \xi} w, w\bs{D \tilde\Sigma D}).
\eer
\noindent {Let $\phi_{\bs Y}(\bs{t})$ denote the characteristic function of $\bs Y$, with $\bs{t} \in {\cal R}^p$. Using the result in \eqref{mixtureALD2}, it follows that:}
\ber \label{chfy0}
\phi_{\bs Y}(\bs{t}) = \mathbb{E}_{W}\left[\mathbb{E}_{Y}[e^{i\bs{t'Y}} \mid W = w]\right] = \int_0^{\infty} \mathbb{E}_{Z}[e^{i\bs{t'}(\bs{\mu} + \bs{D\tilde\xi}w + \sqrt{w} \bs{D \tilde\Sigma}^\frac{1}{2} \bs{Z})} \mid W = w] e^{-w} dw.
\eer
Now, using the conditional distribution of $\bs{Y}$ given $W$ in \eqref{mixturecondALD2}, we have that: 
$$\mathbb{E}_{Z}[e^{i\bs{t'}(\bs{\mu} + \bs{D\tilde\xi}w + \sqrt{w} \bs{D \tilde\Sigma}^\frac{1}{2} \bs{Z})} \mid W = w] = e^{i\bs{t' \mu} + i\bs{t'D\tilde\xi}w -\frac{w}{2} \bs{t'D\tilde\Sigma Dt}}.$$ 

\noindent Substituting this result into \eqref{chfy0} yields:
\ber
\phi_{\bs Y}(\bs{t})&=& e^{i\bs{t' \mu}} \int_0^{\infty} e^{-w(1+ \frac{1}{2} \bs{t'D\tilde\Sigma Dt} -i\bs{t'D \tilde\xi} )} dw.
\eer
\noindent Finally,  integrating over $W$, we obtain:
\ber \label{chfy}
\phi_{\bs Y}(\bs{t})&=& e^{i\bs{t' \mu}} \left(1+ \frac{1}{2} \bs{t'D\tilde\Sigma Dt} -i\bs{t'D \tilde\xi} \right)^{-1}.
\eer

\noindent Now, let $\bs b = (b_1,...,b_p)' \in {\cal R}^p$ be a $p \times 1$ vector such that $\bs b \neq \bs{0}_p$ and consider  a new random variable $Y^{\bs b} = \sum_{j=1}^p b_j Y_j = {\bs b}' {\bs Y}$, having  characteristic function $\phi_{Y^{\bs {b}}}(z)$, with $z \in {\cal R}$. Notice that $Y^{\bs b}$ is a  linear transformation of the marginals $Y_1, \dots, Y_p$. Therefore,  the relation $\phi_{Y^{\bs {b}}}(z) = \phi_{\bs {b' Y}}(z) = \phi_{\bs Y}(\bs{b}z)$ holds, since:

\ber \label{chfyb}
\phi_{Y^{\bs {b}}}(z)&=& e^{iz\bs{b'\mu}} \left(1+ \frac{1}{2} z^2\bs{b'D\tilde\Sigma Db} -iz\bs{b'D \tilde\xi} \right)^{-1}.
\eer

The characteristic function in \eqref{chfyb} resembles the characteristic function of the AL univariate distribution discussed in \cite{yu2001bayesian} and \cite{kozumi2011gibbs} where $\mu^\star, \tau^\star, \delta^\star$ are the scale, skewness and scale parameters, respectively. Therefore, the characteristic function of $Y^{\bs {b}}$ in \eqref{chfyb} can be rewritten as the characteristic function of a univariate AL distribution with parameters:

\ber \label{chfyb1}
\mu^\star = \bs{b'\mu} \mbox{,} \quad  \tau^\star = \frac{1}{2}\left(1-\frac{\bs{b'D\tilde\xi}}{\sqrt{2(\bs{b'D\tilde\Sigma Db}) + (\bs{b'D\tilde\xi})^2}}\right)  \quad  \mbox{and}  \quad \delta^\star=\frac{(\bs{b'D\tilde\Sigma Db})}{2{\sqrt{2(\bs{b'D\tilde\Sigma Db}) + (\bs{b'D\tilde\xi})^2}}}. \nonumber \\[-0.01in]
\eer 
\begin{flushright}
\qedsymbol
\end{flushright}

In conclusion, we obtain that $Y^{\bs b} \sim \mbox{AL} (\mu^\star, \tau^\star, \delta^\star)$ and  $\mathbb{P}(Y^{\bs{b}} < \mu^{\star}) = \tau^{\star}$. Consequently, the parameter $\tau^\star$ controls the probability assigned to each side of $Y^{\bs{b}}$ and $\mu^\star$ is the corresponding quantile at level $\tau^\star$. Notice that the denominator $2(\bs{b'D\tilde\Sigma Db}) + (\bs{b'D\tilde\xi})^2$ in \eqref{chfyb1} is well defined on the positive real line since $(\bs{b'D\tilde\xi})^2 \geq 0$ and $2(\bs{b'D\tilde\Sigma Db}) > 0$ because $\bs {\tilde \Sigma}$ is a positive definite matrix. Furthermore, when $\bs{\tau} = [0.5,0.5,...,0.5]$,  we have that $\bs{b'D\tilde\xi}=0$ and \eqref{chfyb1} simplifies to $\tau^{\star} = 0.5$ and $\delta^\star = \frac{\sqrt{\bs{b'D\tilde\Sigma Db}}}{2 \sqrt{2}}$,  which implies that the distribution of $Y^{\bs b}$ is symmetric around $\mu^\star$. 

}

%
%
%
%



\clearpage
\section*{Appendix B. Simulation study}\label{appB}
In this Appendix we conduct a simulation study to evaluate the finite sample properties of the proposed method and its ability to jointly estimate the pair (VaR, ES) for multiple correlated assets. This simulation exercise addresses the following issues. First, we consider different distributional choices for the error term to investigate the behaviour of the model in the presence of non-Gaussian errors. Second, we evaluate the bias and accuracy of the ML estimators when the interest of the research is focused upon the lower tails of the distribution. Finally, we inspect the impact of dimensionality on both the estimated parameters and the computational burden of the optimization routine. 

In the first experiment, we consider a sample size of $T=1500$ and set $p=3$. \textcolor{black}{The observations are simulated using the following data generating process:
\begin{equation}\label{data_gen}
\bs Y_t = {\cal Q}_{\bs{Y}_{t}}(\bs \tau|\mathcal{F}_{t-1}) + \bs \epsilon_t, \quad \quad t=1,2,...,T,
\end{equation}
where ${\cal Q}_{\bs{Y}_{t}}(\bs \tau|\mathcal{F}_{t-1})$ is} generated according to the three different CAViaR specifications described in \eqref{SAV}-\eqref{IG}. \textcolor{black}{For the} ES component we adopted both the multiplicative factor specification in \eqref{deltaES} and the AR formulation in \eqref{ARES1}-\eqref{ARES2}. Following \cite{petrella2018joint}, two different simulation scenarios are considered for the error terms \textcolor{black}{$\bs \epsilon_t$ in \eqref{data_gen}}:
\begin{enumerate}[label=]
\item (i) a multivariate Normal distribution ($\mathcal{N}_3)$ with zero mean and variance-covariance matrix equal to $\bs{D}_t \bs{\tilde \Sigma} \bs{D}_t$, that is $\bs \epsilon_{it} \sim \mathcal{N}_3(\bs 0, \bs{D}_t \bs{\tilde \Sigma} \bs{D}_t)$;
\item (ii) a multivariate Student-t distribution ($\mathcal{T}_3$) with 5 degrees of freedom, scale parameter $\bs{D}_t \bs{\tilde \Sigma} \bs{D}_t$ and non centrality parameter equal to $\bs D_t\bs {\tilde\xi}$, that is $\bs \epsilon_{it} \sim \mathcal{T}_3 (5, \bs D_t\bs {\tilde\xi}, \bs{D}_t \bs{\tilde \Sigma} \bs{D}_t)$.
\end{enumerate}
The true values of the CAViaR model and the ES dynamics are calibrated using the real data in the empirical application. Specifically, we set
\begin{equation*}\label{betatrue}
\boldsymbol{\omega} = [-0.20, -0.12, -0.24]', \quad \boldsymbol{\eta} = [0.85, 0.70, 0.60]', \quad \boldsymbol{\beta}_1 = [-0.10, -0.05, -0.20]', \quad \boldsymbol{\beta}_2 = [0.05, 0.10, 0.20]'
\end{equation*}
and 
\begin{equation*}\label{gammatrue}
\boldsymbol{\gamma}_0 = [-1.1, -1.5, -1.3]', \quad \boldsymbol{\gamma}_1 = [0.05, 0.10, 0.02]', \quad \boldsymbol{\gamma}_2 = [0.12, 0.05, 0.20]' \quad \textnormal{and} \quad \boldsymbol{\gamma}_3 = [0.80, 0.70, 0.60]'.
\end{equation*}
For the CAViaR-IG dynamic, each element of the vector $\boldsymbol{\omega}$ is considered in absolute value to guarantee that the autoregressive process in \eqref{IG} is well-defined.
Finally, we set $\bs \Psi = \bqmatrix  1 & 0.3 & 0.7 \\ 0.3 & 1 & 0.5 \\ 0.7 & 0.5 & 1 \eqmatrix$.
\\
Since we are interested in evaluating the downside risk, we analyze three different quantile vectors, namely $\bs {\tau} = [0.1, 0.1, 0.1]$, $\bs {\tau} = [0.05, 0.05, 0.05]$ and $\bs {\tau} = [0.01, 0.01, 0.01]$. For each model, we carry out $B=250$ Monte Carlo replications and report the percentage relative bias (Bias\%) and the Root Mean Square Error (RMSE), averaged across the B simulations. 
Tables \ref{tab:s1} and \ref{tab:s2} report the results for all the parameters $\boldsymbol{\omega}, \boldsymbol{\eta}$ and $\boldsymbol{\beta}$ of the three CAViaR specifications. As can be noted, our estimation method is able to recover the true CAViaR specifications under both the $\mathcal{N}_3$- and $\mathcal{T}_3$- scenarios, and for both the considered ES dynamics. Indeed, both the Bias\% and the RMSE remain reasonably small under all the different scenarios even though, as expected, their values tend to slightly increase as the quantile level becomes more extreme (due to the reduced information available at the tails of the distribution) and when we consider a heavy-tailed distribution ($\mathcal{T}_3$-scenario).
 To computationally evaluate the speed of convergence of the EM algorithm, in the last row of each panel we also report the median number of iterations and CPU Time (in seconds) required by the implemented \texttt{R} code using an Intel Xeon E5-2609 2.40GHz processor. Running times range from $9.613$ seconds for the simplest SAV specification with a constant multiplicative factor specified for the ES component, up to $47.242$ seconds for the most complex AS model with autoregressive ES component, confirming the practical feasibility of our optimization algorithm.\\

Finally, to evaluate the impact of dimensionality on the optimization routine, we considered the same simulation experiment \textcolor{black}{in \eqref{data_gen}} with $T = 1500$ and $\bs \tau = [0.1,0.1,...,0.1]$, and let $p = \{3,5,10,12\}$ grow. Specifically, for each value of $p$, Tables \ref{tab:s1a} and \ref{tab:s2b} report the Bias\% and RMSE of the parameter $\theta = \mid \mid \boldsymbol{\theta} \mid \mid$ over 100 Monte Carlo replications, where $\boldsymbol{\theta} = [\boldsymbol{\omega}, \boldsymbol{\eta}, \boldsymbol{\beta}]'$ and $\mid \mid \cdot \mid \mid$ denote the $\ell_1$ norm of $\boldsymbol{\theta}$. As before, we also report the median number of iterations and CPU Time (in seconds) needed to fit the model. As one can easily see, the Bias\% remains relatively constant regardless of the value of $p$, even though, as expected, the RMSE slightly increases with the dimensionality of the problem.
 
\newgeometry{left=05mm,right=05mm}
\begin{table}
\centering
 \smallskip 
 \resizebox{1.0\columnwidth}{!}{
\begin{tabular}{lccccccccc}
\hline
CAViaR & \multicolumn{3}{c}{SAV} & \multicolumn{3}{c}{AS} & \multicolumn{3}{c}{IG}\\\cmidrule(r){2-4}\cmidrule(r){5-7}\cmidrule(r){8-10}
$\bs {\tau}$ & $[0.1, 0.1, 0.1]$ & $[0.05, 0.05, 0.05]$ & $[0.01, 0.01, 0.01]$ & $[0.1, 0.1, 0.1]$ & $[0.05, 0.05, 0.05]$ & $[0.01, 0.01, 0.01]$ & $[0.1, 0.1, 0.1]$ & $[0.05, 0.05, 0.05]$ & $[0.01, 0.01, 0.01]$ \\
\hline
\multicolumn{10}{l}{Panel A: $\bs \epsilon_{it} \sim \mathcal{N}_3(\bs 0, \bs{D}_t \bs{\tilde \Sigma} \bs{D}_t)$}\\
$\omega_1$   & $-0.120 \; (0.027)$ & $0.758 \; (0.047)$   & $3.874 \; (0.149)$  & $-1.621 \; (0.036)$ & $-2.048 \; (0.050)$ & $-3.472 \; (0.151)$ & $-1.494 \; (0.031)$ & $2.520 \; (0.044)$ & $3.749 \; (0.156)$ \\
$\omega_2$   & $-1.020 \; (0.037)$ & $-3.119 \; (0.052)$  & $2.057 \; (0.145)$  & $-1.265 \; (0.030)$ & $-3.738 \; (0.054)$ & $-4.086 \; (0.169)$ & $-1.334 \; (0.036)$ & $1.558 \; (0.053)$ & $3.622 \; (0.142)$ \\
$\omega_3$   & $-1.371 \; (0.028)$ & $-2.245 \; (0.053)$  & $3.900 \; (0.162)$ & $-1.454 \; (0.035)$ & $-2.617 \; (0.063)$ & $-3.238 \; (0.147)$ & $-1.195 \; (0.033)$ & $2.751 \; (0.048)$  & $3.964 \; (0.148)$ \\
$\eta_1$     & $1.184 \; (0.025)$  & $1.337 \; (0.057)$   & $-1.528 \; (0.153)$ & $2.241 \; (0.030)$  & $2.379 \; (0.049)$  & $-3.185 \; (0.155)$ & $0.548 \; (0.048)$  & $-2.690 \; (0.059)$ & $-2.941 \; (0.159)$ \\
$\eta_2$     & $1.228 \; (0.035)$  & $1.753 \; (0.041)$   & $-2.537 \; (0.143)$ & $-0.179 \; (0.034)$ & $-1.855 \; (0.062)$ & $1.992 \; (0.199)$  & $1.794 \; (0.039)$  & $-2.314 \; (0.041)$ & $-3.570 \; (0.144)$ \\
$\eta_3$     & $1.560 \; (0.027)$  & $2.472 \; (0.036)$   & $-1.132 \; (0.174)$ & $1.830 \; (0.041)$  & $2.991 \; (0.062)$  & $4.230 \; (0.178)$  & $1.981 \; (0.031)$  & $2.220 \; (0.054)$  & $2.781 \; (0.153)$  \\
$\beta_{11}$ & $-1.573 \; (0.031)$ & $-2.688 \; (0.038)$ & $3.971 \; (0.155)$  & $-1.409 \; (0.035)$ & $-3.308 \; (0.056)$ & $-3.366 \; (0.168)$ & $-0.141 \; (0.040)$ & $2.704 \; (0.041)$ & $3.497 \; (0.162)$ \\
$\beta_{12}$ & $-0.841 \; (0.033)$ & $-1.521 \; (0.060)$ & $2.459 \; (0.140)$ & $-0.517 \; (0.031)$ & $-0.797 \; (0.045)$ & $1.871 \; (0.159)$  & $1.427 \; (0.045)$  & $1.981 \; (0.062)$ & $2.082 \; (0.185)$ \\
$\beta_{13}$ & $-1.799 \; (0.045)$ & $-2.066 \; (0.057)$ & $3.211 \; (0.165)$ & $-1.439 \; (0.039)$ & $-1.937 \; (0.055)$ & $3.540 \; (0.167)$  & $1.899 \; (0.041)$  & $2.853 \; (0.073)$ & $3.198 \; (0.159)$ \\
$\beta_{21}$ & & & & $1.862 \; (0.033)$  & $2.008 \; (0.046)$  & $5.335 \; (0.147)$  \\
$\beta_{22}$ & & & & $-0.187 \; (0.039)$ & $1.705 \; (0.054)$  & $4.059 \; (0.161)$ \\
$\beta_{23}$ & & & & $0.634 \; (0.039)$  & $1.089 \; (0.049)$  & $3.899 \; (0.158)$   \\
Iterations & 6 & 8 & 12 & 8 & 14 & 27 & 6 & 9 & 14 \\ 
CPU Time & 9.613 & 11.824 & 17.090 & 11.498 & 22.830 & 33.933 & 9.669 & 12.816 & 18.159 \\ 
\hline
\multicolumn{10}{l}{Panel B: $\bs \epsilon_{it} \sim \mathcal{T}_3 (5, \bs D_t\bs {\tilde\xi}, \bs{D}_t \bs{\tilde \Sigma} \bs{D}_t)$}\\
$\omega_1$   & $-1.196 \; (0.039)$ & $-2.393 \; (0.063)$ & $2.996 \; (0.156)$ & $-0.196 \; (0.049)$ & $-2.898 \; (0.051)$ & $-2.393 \; (0.168)$ & $-1.291 \; (0.029)$ & $1.996 \; (0.056)$ & $3.560 \; (0.163)$ \\
$\omega_2$   & $-1.833 \; (0.045)$ & $-1.572 \; (0.059)$ & $1.385 \; (0.162)$ & $-2.833 \; (0.035)$ & $-2.380 \; (0.052)$ & $-4.572 \; (0.179)$ & $-1.533 \; (0.034)$ & $1.385 \; (0.052)$ & $2.608 \; (0.160)$ \\
$\omega_3$   & $-1.397 \; (0.036)$ & $-1.297 \; (0.041)$ & $3.736 \; (0.168)$ & $-2.397 \; (0.046)$ & $-2.120 \; (0.061)$ & $-4.297 \; (0.151)$ & $-1.558 \; (0.032)$ & $1.736 \; (0.068)$ & $1.825 \; (0.169)$ \\
$\eta_1$     & $2.699 \; (0.051)$  & $0.292 \; (0.067)$  & $-2.097 \; (0.170)$ & $2.699 \; (0.041)$  & $2.266 \; (0.057)$  & $2.292 \; (0.167)$  & $1.132 \; (0.027)$  & $-2.097 \; (0.050)$ & $-1.940 \; (0.164)$ \\
$\eta_2$     & $1.050 \; (0.048)$  & $1.849 \; (0.068)$  & $-4.367 \; (0.177)$ & $0.050 \; (0.038)$  & $1.421 \; (0.047)$  & $1.849 \; (0.198)$  & $1.805 \; (0.037)$  & $-2.367 \; (0.047)$ & $-3.034 \; (0.153)$ \\
$\eta_3$     & $1.508 \; (0.042)$  & $2.455 \; (0.061)$  & $3.581 \; (0.167)$  & $1.508 \; (0.032)$  & $1.783 \; (0.049)$ & $3.455 \; (0.151)$  & $1.335 \; (0.036)$  & $1.581 \; (0.057)$  & $2.243 \; (0.140)$  \\
$\beta_{11}$ & $-1.718 \; (0.030)$ & $-0.402 \; (0.050)$ & $2.688 \; (0.154)$ & $-2.718 \; (0.049)$ & $-3.440 \; (0.068)$ & $-3.702 \; (0.160)$ & $-1.908 \; (0.036)$ & $2.688 \; (0.054)$ & $1.954 \; (0.162)$ \\
$\beta_{12}$ & $-1.615 \; (0.048)$ & $-0.402 \; (0.052)$ & $3.189 \; (0.163)$ & $-1.615 \; (0.038)$ & $-2.691 \; (0.064)$ & $-2.402 \; (0.147)$ & $-0.940 \; (0.034)$ & $1.189 \; (0.053)$ & $2.514 \; (0.189)$ \\
$\beta_{13}$ & $-2.557 \; (0.037)$ & $1.942 \; (0.049)$  & $3.087 \; (0.180)$ & $-2.157 \; (0.047)$ & $-2.542 \; (0.060)$ & $1.942 \; (0.199)$  & $1.769 \; (0.041)$  & $2.087 \; (0.080)$ & $3.941 \; (0.170)$ \\
$\beta_{21}$ & & & & $1.371 \; (0.059)$  & $1.057 \; (0.056)$  & $3.782 \; (0.170)$  \\
$\beta_{22}$ & & & & $-0.803 \; (0.040)$ & $-1.690 \; (0.052)$ & $1.166 \; (0.166)$  \\
$\beta_{23}$ & & & & $0.489 \; (0.039)$  & $1.607 \; (0.065)$  & $2.210 \; (0.163)$  \\  
Iterations & 7 & 10 & 15 & 11 & 16 & 28 & 7 & 11 & 16  \\ 
CPU Time & 9.617 & 12.799 & 18.141 & 11.691 & 22.983 & 34.631 & 9.628 & 13.812 & 18.385 \\
\hline
\end{tabular}}
\caption{Bias\% and RMSE (in brackets) of point estimates for $\boldsymbol{\omega}, \boldsymbol{\eta}$ and $\boldsymbol{\beta}$ of the three CAViaR specifications in \eqref{SAV}-\eqref{IG} with the ES modeled as a multiple of VaR as in \eqref{deltaES}, under the $\mathcal{N}_3$- and $\mathcal{T}_3$- scenarios. The last two rows of each Panel show the median number of iterations and CPU Time (in seconds) required to fit the model using a single run of the EM algorithm.}\label{tab:s1}
\end{table}

\begin{table}
\centering
 \smallskip 
 \resizebox{1.0\columnwidth}{!}{
\begin{tabular}{lccccccccc}
\hline
CAViaR & \multicolumn{3}{c}{SAV} & \multicolumn{3}{c}{AS} & \multicolumn{3}{c}{IG}\\\cmidrule(r){2-4}\cmidrule(r){5-7}\cmidrule(r){8-10}
$\bs {\tau}$ & $[0.1, 0.1, 0.1]$ & $[0.05, 0.05, 0.05]$ & $[0.01, 0.01, 0.01]$ & $[0.1, 0.1, 0.1]$ & $[0.05, 0.05, 0.05]$ & $[0.01, 0.01, 0.01]$ & $[0.1, 0.1, 0.1]$ & $[0.05, 0.05, 0.05]$ & $[0.01, 0.01, 0.01]$ \\
\hline
\multicolumn{10}{l}{Panel A: $\bs \epsilon_{it} \sim \mathcal{N}_3(\bs 0, \bs{D}_t \bs{\tilde \Sigma} \bs{D}_t)$}\\
$\omega_1$ & $-0.688 \; (0.037)$  & $-2.193 \;  (0.094)$    & $-3.853 \; (0.155)$  & $-2.266 \;  (0.060)$    & $2.609 \; (0.156)$   & $2.245 \;  (0.196)$ & $0.505 \; (0.059)$   & $-2.726 \; (0.075)$  & $2.522 \; (0.150)$  \\
$\omega_2$ & $-0.413 \; (0.037)$  & $-2.936 \;  (0.053)$    & $-3.462 \; (0.165)$  & $-1.189 \;  (0.065)$    & $2.139 \; (0.055)$   & $3.807 \;  (0.191)$ & $-1.986 \; (0.077)$  & $-2.309 \; (0.084)$  & $3.130 \; (0.162)$  \\
$\omega_3$ & $-1.796 \; (0.055)$  & $-1.046 \;  (0.063)$   & $-3.785 \; (0.145)$  & $-1.949 \;  (0.057)$    & $1.749 \; (0.073)$  & $2.424 \;  (0.165)$ & $-1.162 \; (0.064)$  & $-4.810 \; (0.121)$  & $3.644 \; (0.153)$  \\
$\eta_1$   & $1.691 \; (0.037)$   & $2.940 \;  (0.061)$     & $2.999 \; (0.140)$   & $0.642 \;  (0.061)$     & $-3.533 \; (0.067)$  & $-2.804 \;  (0.199)$ & $1.430 \; (0.040)$   & $-2.116 \; (0.074)$  & $-3.672 \; (0.162)$  \\
$\eta_2$   & $1.884 \; (0.043)$   & $-2.023 \;  (0.050)$   & $-2.133 \; (0.177)$  & $-1.217 \;  (0.052)$    & $-2.763 \; (0.091)$  & $-4.777 \;  (0.140)$ & $-1.813 \; (0.054)$  & $-1.854 \; (0.119)$  & $-3.088 \; (0.155)$ \\
$\eta_3$   & $1.018 \; (0.065)$   & $-1.174 \;  (0.057)$    & $3.017 \; (0.162)$   & $-3.500 \;  (0.065)$    & $-2.376 \; (0.045)$  & $-2.197 \;  (0.150)$ & $1.158 \; (0.081)$   & $-2.133 \; (0.136)$  & $-2.316 \; (0.167)$  \\
$\beta_{11}$ & $-1.850 \; (0.048)$  & $-3.514 \;  (0.045)$    & $3.345 \; (0.133)$  & $1.850 \;  (0.064)$    & $1.395 \; (0.053)$  & $4.856 \;  (0.176)$ & $-2.688 \; (0.060)$  & $1.699 \; (0.102)$  & $2.803 \; (0.189)$  \\
$\beta_{12}$ & $-1.609 \; (0.041)$  & $-2.701 \;  (0.057)$   & $2.465 \; (0.186)$  & $2.602 \;  (0.066)$    & $3.041 \; (0.091)$  & $3.230 \;  (0.141)$ & $-2.357 \; (0.074)$ & $-2.640 \; (0.128)$ & $2.921 \; (0.182)$  \\
$\beta_{13}$ & $-1.224 \; (0.038)$ & $-1.291 \;  (0.045)$   & $2.808 \; (0.164)$  & $1.659 \;  (0.063)$    & $2.407 \; (0.067)$  & $4.289 \;  (0.119)$ & $-1.928 \; (0.065)$ & $-2.191 \; (0.117)$ & $4.053 \; (0.143)$  \\
$\beta_{21}$ & & & & $1.320 \;  (0.058)$     & $-2.763 \; (0.141)$ & $-2.013 \;  (0.193)$  \\
$\beta_{22}$ & & & & $1.599 \;  (0.069)$    & $-2.484 \; (0.121)$ & $-3.276 \; (0.162)$ \\
$\beta_{23}$ & & & & $1.082 \;  (0.060)$    & $-2.179 \; (0.118)$  & $3.936 \;  (0.174)$  \\
Iterations & 12 & 14 & 19 & 15 & 23 & 35 & 10 & 15 & 18 \\ 
CPU Time & 13.791 & 20.739 & 25.738 & 16.112 & 22.956 & 45.946 & 13.796 & 20.781 & 25.784 \\ 
\hline
\multicolumn{10}{l}{Panel B: $\bs \epsilon_{it} \sim \mathcal{T}_3 (5, \bs D_t\bs {\tilde\xi}, \bs{D}_t \bs{\tilde \Sigma} \bs{D}_t)$}\\
$\omega_1$   & $-1.687 \; (0.049)$  & $-3.265 \; (0.121)$  & $-3.792 \; (0.151)$  & $-2.793 \; (0.037)$  & $2.309 \; (0.126)$   & $3.497 \; (0.215)$ & $-1.006 \; (0.038)$  & $-2.278 \; (0.117)$ & $3.580 \; (0.142)$  \\
$\omega_2$   & $-2.891 \; (0.078)$  & $-2.992 \; (0.084)$  & $-4.143 \; (0.189)$  & $-1.244 \; (0.048)$ & $4.041 \; (0.121)$   & $4.667 \; (0.223)$ & $-2.308 \; (0.069)$  & $-3.344 \; (0.102)$  & $2.875 \; (0.127)$  \\
$\omega_3$   & $-1.169 \; (0.087)$ & $-2.719 \; (0.128)$ & $-4.852 \; (0.156)$  & $-1.529 \; (0.070)$ & $1.858 \; (0.093)$  & $2.045 \; (0.216)$ & $-1.845 \; (0.044)$ & $-2.442 \; (0.125)$ & $4.644 \; (0.194)$  \\
$\eta_1$     & $2.834 \; (0.038)$   & $3.721 \; (0.096)$   & $-3.089 \; (0.157)$  & $1.081 \; (0.057)$   & $2.490 \; (0.171)$   & $-3.874 \; (0.190)$ & $1.595 \; (0.051)$   & $2.387 \; (0.093)$   & $-4.524 \; (0.146)$  \\
$\eta_2$     & $-1.345 \; (0.062)$  & $-2.788 \; (0.166)$ & $-4.423 \; (0.154)$  & $-1.449 \; (0.063)$  & $-2.036 \; (0.176)$  & $-3.810 \; (0.215)$ & $0.404 \; (0.059)$   & $-2.002 \; (0.086)$  & $-4.258 \; (0.189)$  \\
$\eta_3$     & $1.642 \; (0.067)$   & $-2.089 \; (0.094)$  & $-2.026 \; (0.144)$  & $-1.104 \; (0.073)$  & $2.716 \; (0.096)$   & $-2.981 \; (0.221)$ & $2.186 \; (0.057)$   & $3.929 \; (0.106)$   & $2.565 \; (0.192)$   \\
$\beta_{11}$ & $-2.171 \; (0.046)$  & $-3.313 \; (0.091)$  & $2.090 \; (0.132)$   & $0.920 \; (0.066)$  & $3.032 \; (0.133)$   & $3.917 \; (0.186)$ & $-1.893 \; (0.040)$ & $-3.994 \; (0.138)$  & $3.613 \; (0.161)$  \\
$\beta_{12}$ & $-2.212 \; (0.050)$  & $-3.605 \; (0.112)$ & $1.394 \; (0.146)$  & $1.261 \; (0.082)$  & $2.959 \; (0.090)$  & $3.883 \; (0.197)$ & $-1.525 \; (0.047)$ & $-2.143 \; (0.104)$ & $3.346 \; (0.181)$ \\
$\beta_{13}$ & $-1.536 \; (0.042)$  & $-2.111 \; (0.085)$ & $3.773 \; (0.129)$  & $1.971 \; (0.082)$  & $3.189 \; (0.145)$  & $2.862 \; (0.184)$ & $-2.807 \; (0.054)$ & $-3.592 \; (0.092)$  & $4.842 \; (0.164)$ \\
$\beta_{21}$ & & & & $1.171 \; (0.051)$   & $1.847 \; (0.137)$ & $3.866 \; (0.177)$ \\
$\beta_{22}$ & & & & $1.229 \; (0.067)$  & $2.621 \; (0.154)$ & $3.195 \; (0.196)$ \\
$\beta_{23}$ & & & & $1.456 \; (0.063)$  & $2.158 \; (0.133)$ & $3.535 \; (0.187)$ \\    
Iterations & 13 & 14 & 20 & 16 & 24 & 37 & 11 & 16 & 20  \\ 
CPU Time & 13.797 & 20.760 & 25.790 & 16.783 & 22.176 & 47.242 & 13.794 & 20.752 & 26.766  \\ 
\hline
\end{tabular}}
\caption{Bias\% and RMSE (in brackets) of point estimates for $\boldsymbol{\omega}, \boldsymbol{\eta}$ and $\boldsymbol{\beta}$ of the three CAViaR specifications in \eqref{SAV}-\eqref{IG} with the AR process for the ES as in \eqref{ARES1}-\eqref{ARES2}, under the $\mathcal{N}_3$- and $\mathcal{T}_3$- scenarios. The last two rows of each Panel show the median number of iterations and CPU Time (in seconds) required to fit the model using a single run of the EM algorithm.}\label{tab:s2}
\end{table}
\restoregeometry

\begin{table}[h]
\centering
 \smallskip 
 \resizebox{1.0\columnwidth}{!}{
\begin{tabular}{lcccccccccccc}
\hline
 & \multicolumn{3}{c}{$p=3$} & \multicolumn{3}{c}{$p=5$} & \multicolumn{3}{c}{$p=10$} & \multicolumn{3}{c}{$p=12$}\\\cmidrule(r){2-4}\cmidrule(r){5-7}\cmidrule(r){8-10}\cmidrule(r){11-13}
CAViaR & SAV & AS & IG & SAV & AS & IG & SAV & AS & IG & SAV & AS & IG\\
\hline
\multicolumn{10}{l}{Panel A: $\bs \epsilon_{it} \sim \mathcal{N}_3(\bs 0, \bs{D}_t \bs{\tilde \Sigma} \bs{D}_t)$}\\
Bias\% & 0.754 & 0.672 & 0.855 & 0.697 & 0.742 & 0.748 & 0.803 & 0.729 & 0.764 & 0.869 & 0.680 & 0.654 \\ 
RMSE & 0.083 & 0.088 & 0.075 & 0.118 & 0.390 & 0.272 & 0.299 & 0.267 & 0.315 & 0.351 & 0.391 & 0.341\\
Iterations & 6 & 9 & 6 & 12 & 14 & 11 & 18 & 27 & 22 & 47 & 55 & 48 \\
CPU Time & 9.812 & 11.746 & 9.756 & 46.207 & 60.240 & 47.248 & 88.980 & 102.840 & 98.040 & 126.540 & 137.880 & 120.660\\
\hline
\multicolumn{10}{l}{Panel B: $\bs \epsilon_{it} \sim \mathcal{T}_3 (5, \bs D_t\bs {\tilde\xi}, \bs{D}_t \bs{\tilde \Sigma} \bs{D}_t)$}\\
Bias\% & 0.837 & 0.809 & 0.785 & 0.793 & 0.884 & 0.895 & 0.734 & 0.831 & 0.702 & 0.893 & 0.708 & 0.929 \\ 
RMSE & 0.095 & 0.085 & 0.083 & 0.207 & 0.375 & 0.321 & 0.372 & 0.396 & 0.380 & 0.362 & 0.379 & 0.442\\
Iterations & 6 & 9 & 7 & 11 & 16 & 12 & 18 & 29 & 23 & 48 & 57 & 50 \\
CPU Time & 9.783 & 11.742 & 9.776 & 47.900 & 64.440 & 48.944 & 84.720 & 98.840 & 96.760 & 130.200 & 164.100 & 155.460 \\
\hline
\end{tabular}}
\caption{Bias\% and RMSE of point estimates of $\theta = \mid \mid \boldsymbol{\theta} \mid \mid$, where $\boldsymbol{\theta} = [\boldsymbol{\omega}, \boldsymbol{\eta}, \boldsymbol{\beta}]'$, for different values of $p$ and for the three CAViaR specifications in \eqref{SAV}-\eqref{IG}, with the ES modeled as a multiple of VaR as in \eqref{deltaES}. Panels A and B refer to the $\mathcal{N}_3$- and $\mathcal{T}_3$- scenarios, respectively, where the last two rows of each panel show the median number of iterations and CPU Time (in seconds) required to fit the model using a single run of the EM algorithm.}\label{tab:s1a}
\end{table}
\restoregeometry

\begin{table}[h]
\centering
 \smallskip 
 \resizebox{1.0\columnwidth}{!}{
\begin{tabular}{lcccccccccccc}
\hline
 & \multicolumn{3}{c}{$p=3$} & \multicolumn{3}{c}{$p=5$} & \multicolumn{3}{c}{$p=10$} & \multicolumn{3}{c}{$p=12$}\\\cmidrule(r){2-4}\cmidrule(r){5-7}\cmidrule(r){8-10}\cmidrule(r){11-13}
CAViaR & SAV & AS & IG & SAV & AS & IG & SAV & AS & IG & SAV & AS & IG\\
\hline
\multicolumn{10}{l}{Panel A: $\bs \epsilon_{it} \sim \mathcal{N}_3(\bs 0, \bs{D}_t \bs{\tilde \Sigma} \bs{D}_t)$}\\
Bias\% & 0.786 & 0.791 & 0.773 & 0.731 & 0.865 & 0.671 & 0.912 & 0.713 & 0.792 & 0.803 & 0.753 & 0.800 \\ 
RMSE & 0.140 & 0.149 & 0.147 & 0.265 & 0.279 & 0.232 & 0.243 & 0.322 & 0.223 & 0.364 & 0.358 & 0.437\\
Iterations & 12 & 13 & 11 & 23 & 24 & 21 & 38 & 37 & 35 & 67 & 75 & 68 \\
CPU Time & 12.617 & 15.799 & 13.141 & 82.140 & 89.400 & 85.400 & 109.680 & 168.240 & 118.320 & 196.620 & 303.360 & 196.540 \\
\hline
\multicolumn{10}{l}{Panel B: $\bs \epsilon_{it} \sim \mathcal{T}_3 (5, \bs D_t\bs {\tilde\xi}, \bs{D}_t \bs{\tilde \Sigma} \bs{D}_t)$}\\
Bias\% & 0.722 & 0.729 & 0.742 & 0.704 & 0.783 & 0.639 & 0.737 & 0.765 & 0.762 & 0.828 & 0.734 & 0.633 \\ 
RMSE & 0.152 & 0.180 & 0.152 & 0.256 & 0.297 & 0.302 & 0.315 & 0.356 & 0.283 & 0.393 & 0.425 & 0.395\\
Iterations & 12 & 14 & 12 & 25 & 23 & 23 & 38 & 39 & 43 & 68 & 77 & 70 \\
CPU Time & 12.681 & 16.824 & 13.090 & 89.880 & 94.920 & 83.652 & 106.260 & 168.360 & 117.960 & 177.420 & 314.840 & 258.620\\
\hline
\end{tabular}}
\caption{Bias\% and RMSE of point estimates of $\theta = \mid \mid \boldsymbol{\theta} \mid \mid$, where $\boldsymbol{\theta} = [\boldsymbol{\omega}, \boldsymbol{\eta}, \boldsymbol{\beta}]'$, for different values of $p$ and for the three CAViaR specifications in \eqref{SAV}-\eqref{IG}, with the AR process for the ES as in \eqref{ARES1}-\eqref{ARES2}. Panels A and B refer to the $\mathcal{N}_3$- and $\mathcal{T}_3$- scenarios, respectively, where the last two rows of each panel show the median number of iterations and CPU Time (in seconds) required to fit the model using a single run of the EM algorithm.}\label{tab:s2b}
\end{table}
\restoregeometry

\clearpage
\bibliographystyle{chicago}

\bibliography{biblio}
\end{document}